\documentclass[journal=jpclcd,manuscript=letter,layout=onecolumn]{achemso}

\SectionNumbersOn
\usepackage[version=3]{mhchem} 
\usepackage{graphicx}
\usepackage{dcolumn}
\usepackage{bm}
\usepackage{xcolor}
\usepackage{amsmath}
\usepackage{longtable}
\usepackage[english]{babel}
\usepackage[utf8]{inputenc}
\usepackage{braket}
\usepackage{float}
\usepackage{tablefootnote}
\usepackage{bibunits}

\DeclareMathOperator{\Tr}{Tr}

\newcommand{\fixme}[1]{\textcolor{black}{#1}}

\makeatletter
\renewcommand{\acs@tocentry@width}{3.25in}
\renewcommand{\acs@tocentry@height}{1.75in}
\makeatother

\author{Torsha Moitra}
\email{torsha.moitra@uniba.sk}
\affiliation{Hylleraas Centre for Quantum Molecular Sciences, Department of Chemistry, UiT The Arctic University of Norway, 9037 Troms{\o}, Norway}
\alsoaffiliation{Department of Physical and Theoretical Chemistry, Faculty of Natural Sciences, Comenius University, 84215 Bratislava, Slovakia}
\author{Lukas Konecny}
\affiliation{Hylleraas Centre for Quantum Molecular Sciences, Department of Chemistry, UiT The Arctic University of Norway, 9037 Troms{\o}, Norway}
\alsoaffiliation{Department of Inorganic Chemistry, Faculty of Natural Sciences, Comenius University, 84215 Bratislava, Slovakia}
\alsoaffiliation{Max Planck Institute for the Structure and Dynamics of Matter, Center for Free Electron Laser Science, Luruper Chaussee 149, 22761 Hamburg, Germany}
\author{Marius Kadek}
\affiliation{Hylleraas Centre for Quantum Molecular Sciences, Department of Chemistry, UiT The Arctic University of Norway, 9037 Troms{\o}, Norway}
\author{Ofer Neufeld}
\affiliation{Technion Israel Institute of Technology, Faculty of Chemistry, Haifa 3200003, Israel}
\author{Angel Rubio}
\email{angel.rubio@mpsd.mpg.de}
\affiliation{Max Planck Institute for the Structure and Dynamics of Matter, Center for Free Electron Laser Science, Luruper Chaussee 149, 22761 Hamburg, Germany}
\alsoaffiliation{Initiative for Computational Catalysis (ICC), The Flatiron Institute, 162 Fifth Avenue, New York, New York 10010, USA}
\author{Michal Repisky}
\email{michal.repisky@uit.no}
\affiliation{Department of Physical and Theoretical Chemistry, Faculty of Natural Sciences, Comenius University, 84215 Bratislava, Slovakia}
\alsoaffiliation{Hylleraas Centre for Quantum Molecular Sciences, Department of Chemistry, UiT The Arctic University of Norway, 9037 Troms{\o}, Norway}

\title[]
  {Light-Induced Persistent Electronic Chirality in Achiral Molecules Probed with \fixme{Time-Resolved Electronic} Circular Dichroism Spectroscopy}

\begin{document}
\begin{tocentry}
\vspace{0.4cm}
\includegraphics[]{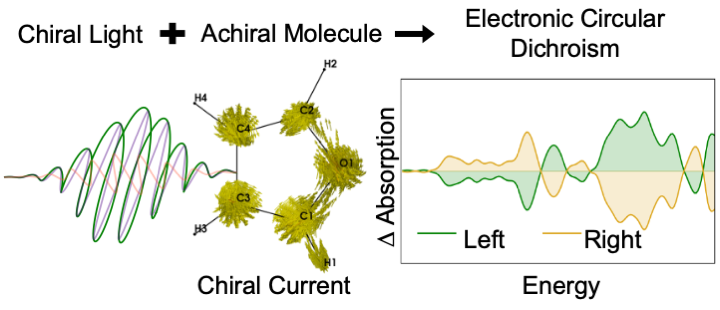}
\end{tocentry}

\begin{abstract}
Chiral systems exhibit unique properties traditionally linked to their asymmetric spatial arrangement. Recently, multiple laser pulses were shown to induce purely electronic chiral states without altering the nuclear configuration. Here, we propose and numerically demonstrate a simpler realization of light-induced electronic chirality that is long-lived and occurs well before the onset of nuclear motion and decoherence. A single monochromatic circularly-polarized laser pulse is shown to induce electronic chiral currents in an oriented achiral molecule. Using state-of-the-art ab initio theory, we analyze this effect and relate the chiral currents to induced magnetic dipole moments, detectable via attosecond \fixme{time-resolved electronic circular dichroism (TR-ECD) spectroscopy, also known as transient absorption ECD}. The resulting chiral electronic wavepacket oscillates rapidly in handedness at harmonics of the pump laser’s carrier frequency, and the currents persist after the pulse ends. We establish a chiral molecular-current analogue to high harmonic generation, and demonstrate attosecond transient chirality control with potential impact on spintronics and reaction dynamics.
\end{abstract}

Chirality is a ubiquitous phenomenon observed in nature associated with the lack of mirror-image superimposability of systems ranging from macroscopic objects to molecules~\cite{1966_AngweChem_Chirality, 2007_book_chirality}. Traditionally, this concept is linked to geometric chirality, where the spatial arrangement of atoms within a molecule creates distinct left and right handedness. Light-induced \emph{nuclear} dynamics operating on femtosecond timescales have shed light on intriguing phenomena where achiral molecules dynamically evolve into geometrically chiral structures via the loss of molecular symmetry\fixme{~\cite{2022_EPJST_psECD-rev, 2022_PCCP_Smirnova_UltrafastChirality-rev, 2022_NatChem_Chergui_Fe-chiral-fsdynamics, 2019_Optica_Chergui_TR-ECD,2017_Science_chiral-ionization-camphor,tikhonov2022pump,leibscher2024quantum}}. However, it was only recently that the concept of chirality has been realized without the intervention of nuclear degrees of freedom by pure \emph{electronic} motion, giving rise to molecular light-induced electronic chirality~\cite{2024_Nat_Wanie_attochirality, 2024_NatComm_Yonggang_attochirality}. Closely intertwined with the concept of electronic chirality are ring currents, which are known to be induced by chiral light and light with orbital angular momentum~\cite{2021_Forbes_OAM}. Helical ring-currents can circulate through molecules \fixme{\cite{2006_JACS_barth_ringcurrents, barth2006periodic, hermann2016multidirectional, 2018_NatPhys_Eckart_ringcurrents, 2019_PRL_Ofer_ringcurrents, 2023_JPCL_Ofer_ringcurrents}} or outside of the molecule \cite{2017_PRL_Pengel_vortices}, breaking all spatial mirror/inversion symmetries of the system in 3D~\cite{2017_PRA_Pengel_3dvortices,2019_NatPhotonics_Smirnova_chirality,2022_PRL_Smirnova_chirality}. In previous works, symmetry breaking has been achieved by using two or more laser pulses, which is intuitively required in order to excite the molecule in a 3D spatial arrangement that breaks all relevant symmetries~\cite{2022_PCCP_Smirnova_UltrafastChirality-rev, 2024_NatRevPhys_Ofer_review,2024_NatComm_Yonggang_attochirality,2019_Ofer_floquetHHG}. 

Experimentally, ultrafast chiral states have been measured on the femtosecond timescale using time-resolved photoemission in atoms \cite{2018_NatPhys_Eckart_ringcurrents}, and time-resolved photoelectron circular dichroism in chiral molecules~\cite{2024_Nat_Wanie_attochirality,2017_Science_chiral-ionization-camphor}. However, to our knowledge no work to date investigated pure attosecond chiral electron dynamics triggered by a single monochromatic light pulse without ionizing the molecule or breaking it apart. Such studies have been performed for magnetism in solids~\cite{2019_Nature_Siegrist_magnetism,2023_npjComputationlMat_Ofer_magnetisation}, but it remains unclear if it is possible to connect attosecond circular-dichroic absorption spectra to chiral currents in molecules. Novel schemes for chiral state control could enable wide-ranging applications such as electronic switches~\cite{2011_Morgen_chiralityflip}, spintronics~\cite{2022_AdvMat_CISS_review,2012_JPCL_Naaman_CISS}, and phase transitions~\cite{2016_bisoyi_ChemRev}.

In this Letter, we study light-induced chirality originating from pure electron dynamics in structurally achiral furan molecule. We show that a single monochromatic circularly-polarized pulse is capable of inducing a chiral non-stationary electronic wavepacket in an oriented sample. This effect can be attributed to the induced ring currents generated by the pump pulse, which evolve dynamically in 3D. Interestingly, the current density is long-lived and the non-stationary chiral state does not instantaneously relax to an achiral state, therefore allowing its detection and monitoring and suggesting a way to generate chiral states that should survive up to few hundred femtoseconds before dephasing. We predict the spectroscopic signature of these chiral wavepackets based on the induced magnetic dipole moments by time-resolved electronic circular dichroism (TR-ECD), also referred to as transient absorption circular dichroism spectroscopy. These moments rapidly oscillate with frequencies corresponding not only to the pump laser’s carrier frequency, but also to its higher-order harmonics that survive long after the laser, and permits attosecond transient chiral states. Our work should therefore pave the way to attosecond chiral state manipulation and readout. 

\begin{figure}[htb!]
    \centering
    \includegraphics[width=0.5\linewidth]{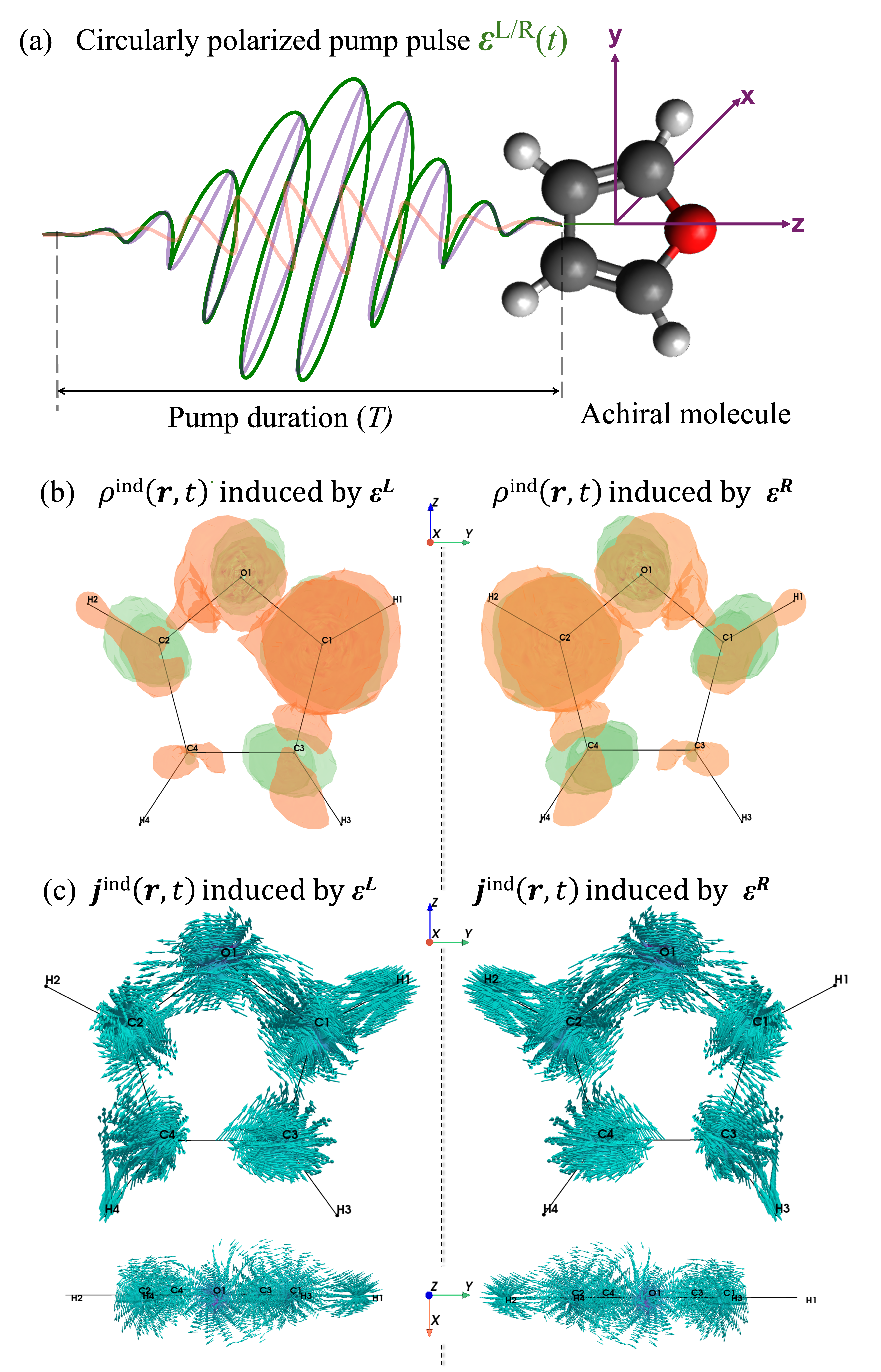}
    \caption{(a) Schematic representation of a circularly-polarized pump pulse
 ${\bm{\mathcal{E}}}^{\text{L/R}}(t)$ in green with $xy$ plane of polarization and
 $z$ propagation, interacting with the achiral furan molecule in $yz$ plane. 
 The chiral induced (b) electron charge density $\rho^{\text{ind}}(\bm{r},t)$ and (c) current density ${\bm{j}}^{\text{ind}}(\bm{r},t)$
 generated by the circularly-polarized light at the end of the pump pulse ($t=T$). 
 Gain and loss of charge density is shown by green and orange color surfaces with isovalue 0.005, respectively. }
    \label{fig:setup-chirality}
\end{figure}
\begin{figure*}[htb!]
    \centering
    \includegraphics[width=\linewidth]{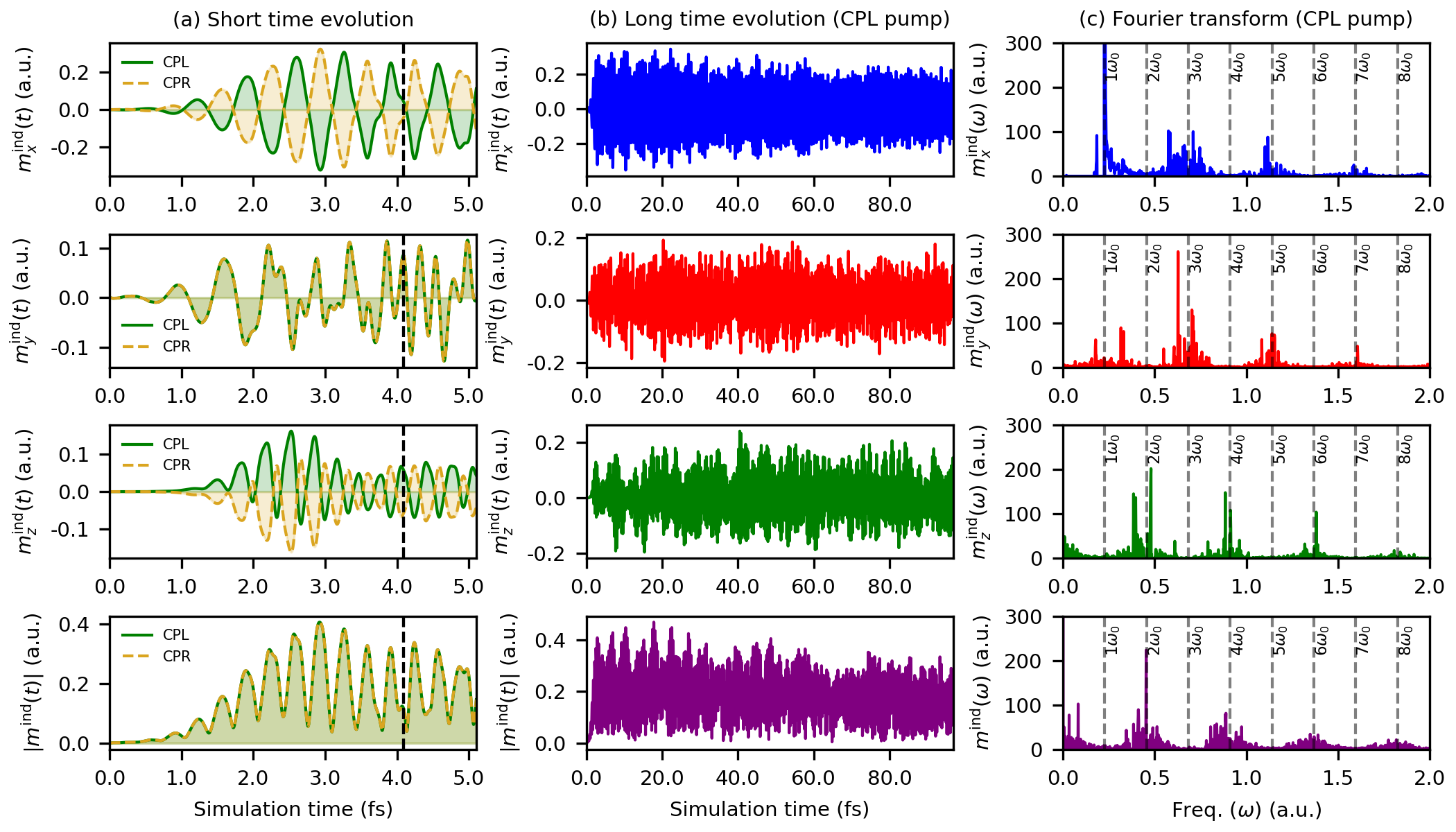}
    \caption{Short time evolution (a) and long time evolution (b) of the magnitude ($|\bm{m}^{\text{ind}}(t)|$) and components of the induced magnetic dipole moment $\mathbf{m}^{\text{ind}}(t)=(m^{\text{ind}}_x(t), m^{\text{ind}}_y(t), m^{\text{ind}}_z(t))$ by CPL and/or CPR pump pulses. The black dashed line in (a) at 4.09 fs marks the end of the pump pulse, after which the electronic wavepacket evolves freely. 
    (c) Fourier transform of the induced magnetic moment obtained from the long time evolution in (b). 
    The harmonic orders of the carrier frequency $\omega_0=0.223$ au are marked by dashed lines.
    Note that the simulation does not include dissipative processes.
    }
    \label{fig:m_t}
\end{figure*}

This study focuses on the furan molecule, which exhibits: (i) an achiral ground-state electronic configuration, (ii) a static electric dipole moment, 
which makes it orientable~\cite{2011_PRL_Fleischer_orientation}, and (iii) is frequently used in studies of other ultrafast phenomena~\cite{2010_JCP_Fuji_furan,2015_CP_Liu_furan,2016_SD_Hua_furan,2024_JACS_Mukamel_TRMCD, 2024_JPCL_Uenishi_furan}. 
Supporting information extends this study to benzene and aniline molecules. 
Figure~\ref{fig:setup-chirality}a illustrates our proposed setup for inducing chiral electron dynamics, where the furan molecule is oriented such that its static electric dipole moment is aligned with the direction of propagation of the CP light (along $z$). The molecule is initially pumped by a chiral, CP left (L) or right (R) laser pulse, whose electric field $\bm{\mathcal{E}}^\text{L/R}(t)$ traces a circular trajectory in the $xy$-plane:
\begin{align}
	\label{eq:pump}
	\bm{\mathcal{E}}^{\text{L/R}} (t) 
         &= 
        \mathcal{E}_0 g(t;t_0,T) [\cos(\omega_0(t-t_0))\bm{x}  
        \mp \sin(\omega_0(t-t_0))\bm{y}],
\end{align}
\fixme{where the negative and positive combinations correspond to left (L) and right (R) circularly polarized light, respectively.}
Here, $g(t;t_0,T)$ is a dimensionless Gaussian envelope centered at $t_0$ and duration $T$, while $\mathcal{E}_0$ and $\omega_0$ are the amplitude and carrier frequency of the monochromatic light pulse, respectively. Additional details about the pump pulse is given in Section~S2. The carrier frequency was tuned to the first bright electronic transition at energy $\hbar\omega_0 = 0.223$~au = 6.07 eV, while $T$ was set to 4.09~fs 
so that the pump pulse populates primarily the first excited state but also encompasses the second excited state at energy $\hbar\omega_1 = 0.294$~au = 7.99 eV. A pump pulse amplitude of $\mathcal{E}_0 = 0.03 $ au corresponding to a peak intensity of  $3.16 \times 10^{13}\text{~W/cm}^2$ is used, which leads to about $3\%$ ground state depopulation at the end of the pump, as shown in Section~S3.

The interaction of an external pulse(s) with the molecule is described from first principles by real-time time-dependent density functional theory (RT-TDDFT) as implemented in the ReSpect program\fixme{~\cite{2020_JCP_ReSpect,repisky2025respect}}. The electronic wavepacket is evolved in time-domain as per the Liouville-von Neumann (LvN) equation of motion, 
\begin{align}
	i \frac{\partial \mathbf{D}(t)}{\partial t} &= [\mathbf{F}(t),  \mathbf{D}(t)] 
\end{align}
where $\mathbf{D}(t)$ and $\mathbf{F}(t)$ are the time-dependent one-electron reduced density matrix and the Fock matrix, respectively. $\mathbf{D}(t)$
represents the state of the system, whereas $\mathbf{F}(t)$ characterizes the molecular system and its interaction with the external electric field within the dipole approximation. Both matrices are represented in molecular orbitals (MOs), where each MO is expressed as a linear combination of Gaussian-type orbitals (GTOs). Here, we employed uncontracted aug-cc-pVTZ~\cite{1989_dunning_basis_1, 1992_dunning_basis_2} GTOs and PBE0 exchange-correlation functional~\cite{PBE0}. We refer the reader to Section~S1 for details on the RT-TDDFT methodology.

In order to get insights into ultrafast electron dynamics we compute the time-dependent induced electron charge
($\rho^{\text{ind}}$) and current (${\bm{j}}^{\text{ind}}$) densities generated by the external pump pulse $\bm{\mathcal{E}}^{\text{L/R}}(t)$. In our formalism, these quantities are defined over ground-state MOs 
$(\phi)$ as, 
\begin{align}
   \rho^{\text{ind}}(\bm{r},t)
   &=
   \Tr{[\mathbf{D}^{\text{ind}}(t)\bm{\Omega}(\bm{r})]}
\\
  j_{k}^{\text{ind}}(\bm{r},t)
  &=
  \Tr{[\mathbf{D}^{\text{ind}}(t)\bm{J}_{k}(\bm{r})]}
  ,\quad k\in x,y,z
\end{align}
where $\bm{\Omega}$ and $\bm{J}_{k}$ are matrices of the charge density and current density operators
\begin{align}
  \Omega_{pq}(\bm{r}) 
  & = 
  \phi_p^\dagger(\bm{r}) \phi_q (\bm{r})
  \\
  J_{k,pq}(\bm{r})
  &=
  -\frac{1}{2}
  \Big( \phi_p^\dagger(\bm{r})\big\{p_{k}\phi_q(\bm{r})\big\} 
     + \big\{p_{k}\phi_p(\bm{r})\big\}^{\dagger} \phi_q(\bm{r})   
  \Big).
\end{align}
The induced density matrix $\mathbf{D}^{\text{ind}}(t)$ is obtained as the difference between $\mathbf{D}(t)$ and the static ground-state $\mathbf{D}(0)$, which evolves on attosecond timescales.

The CPL and CPR pump pulses generate mirror-imaged induced charge and current densities, mimicking molecular enantiomers, as shown in Figs.~\ref{fig:setup-chirality}b and \ref{fig:setup-chirality}c, respectively. The induced current density has both in-plane and out-of-plane contributions, as shown by the top view and side view in Fig.~\ref{fig:setup-chirality}c. See supporting information video for time-evolution of the induced charge and current densities. Most importantly, the induced current density gives rise to a corresponding induced magnetic dipole moment in the system, evaluated as
\begin{equation}
  \bm{m}^{\text{ind}}(t)
  =
  \frac{1}{2} \int~d^3\bm{r}~\bm{r} \times {\bm{j}}^{\text{ind}}(\bm{r},t).
\end{equation}
Figure~\ref{fig:m_t}a displays all components of the induced magnetic moment generated by CPL and CPR pump pulses in green and yellow, respectively. Notably, these induced moments persist well beyond the end of the pump pulse -- indicated by the black dashed line in Fig.~\ref{fig:m_t}a -- and exhibit periodic sign reversals. Furthermore, the overall magnitude of the induced magnetic moment $|\bm{m}^{\text{ind}}(t)|$ remains identical for both laser polarizations. A more detailed inspection of the individual components shows that $m^{\text{ind}}_x(t)$ and $m^{\text{ind}}_z(t)$ have opposite sign for CPL and CPR induced wavepacket, while $m^{\text{ind}}_y(t)$ has the same sign for both. This enantiomeric relationship is preserved during and after the pump pulse. Given the relative orientation of our pump laser with respect to the furan molecule, $m^{\text{ind}}_x(t)$ has the largest magnitude and can be attributed to the out-of-plane enantiomeric dynamics governed by $\pi$-bonding orbitals. A similar enantiomeric behavior is also observed for in-plane $m^{\text{ind}}_z(t)$ component, but with lower magnitude. \fixme{It is important to note that the alternating ring currents and magnetic moments discussed here are induced by a single circularly polarized laser pulse that excites at least two non-degenerate states, as demonstrated in Section~S2. In contrast, for molecules belonging to non-Abelian point groups, which possess sets of degenerate states, such laser pulses can induce unidirectional electronic ring currents without any reversals~\cite{2006_JACS_barth_ringcurrents,barth2006periodic,hermann2016multidirectional}, provided that the carrier frequencies of the laser pulses are resonant with the excitations of selected doubly degenerate states.
}

Fig.~\ref{fig:m_t}b presents the oscillations in the induced magnetic moment for the CPL induced wavepacket propagated for a long timescale of up to 100 fs. We would like to stress here that our simulations incorporate only pure electron dynamics, and have no external decoherence effects (nuclear motion, solvent effect, etc.), which would gain prominence at such large timescales\fixme{\cite{liu2024cpc}}. Nevertheless, even in this idealized scenario, the oscillations of the induced magnetic moment do not exhibit a single dominant frequency. A Fourier transform of the time-domain signal shows contribution from multiple harmonic orders of the laser pulse carrier frequency ($\omega_0$), marked by dashed lines in Figure~\ref{fig:m_t}c. 
Therefore, it is impossible to define a single characteristic frequency for the flipping of the chiral state from one form to the other. 
The presence of higher harmonics also indicates that the induced electronic chirality opens up a route to coherent manipulation of ultrafast chiral states. The observation of multiple harmonic orders in the induced magnetic dipole moment suggests a light-matter mechanism analogous to high harmonic generation (HHG)~\cite{2015_Bharadwaj_NatPhys_cHHG,2019_Ofer_PRX_HHG_chiral,2022_Ayuso_Chiral_HHG}, but where dynamics involved multiple harmonics even in much longer timescales, and in molecular magnetic moments rather than the typical dipole response that emits photons. 
\fixme{The observed higher-order harmonics in the chiral current tune the speed with which the molecular handedness can ultimately be controlled, and here they should be interpreted solely as signatures of intense non-linear responses rather than molecular ionizations.}
Moreover, the observation that even harmonic orders appear in $m^\text{ind}_y(\omega)$, while odd harmonic orders are present in $m^\text{ind}_x(\omega)$ and $m^\text{ind}_z(\omega)$, suggests a symmetry-guided selection rule at play, possibly due to the parity of the chiral wavepacket~\cite{2019_Ofer_floquetHHG}.
The corresponding induced electric dipole moment and its Fourier transform is shown in Section~S4. This rich harmonic content not only provides a spectral fingerprint of the underlying chiral dynamics and highlights the complex interplay between electronic motion and the system's symmetry.
Overall, these results suggest a route towards obtaining attosecond chiral transients even in the absence of using attosecond circularly-polarized pulses, which are very difficult to generate experimentally~\cite{2018_NatPhotonics_polarization}.

\begin{figure}
    \centering
    \includegraphics[width=0.5\linewidth]{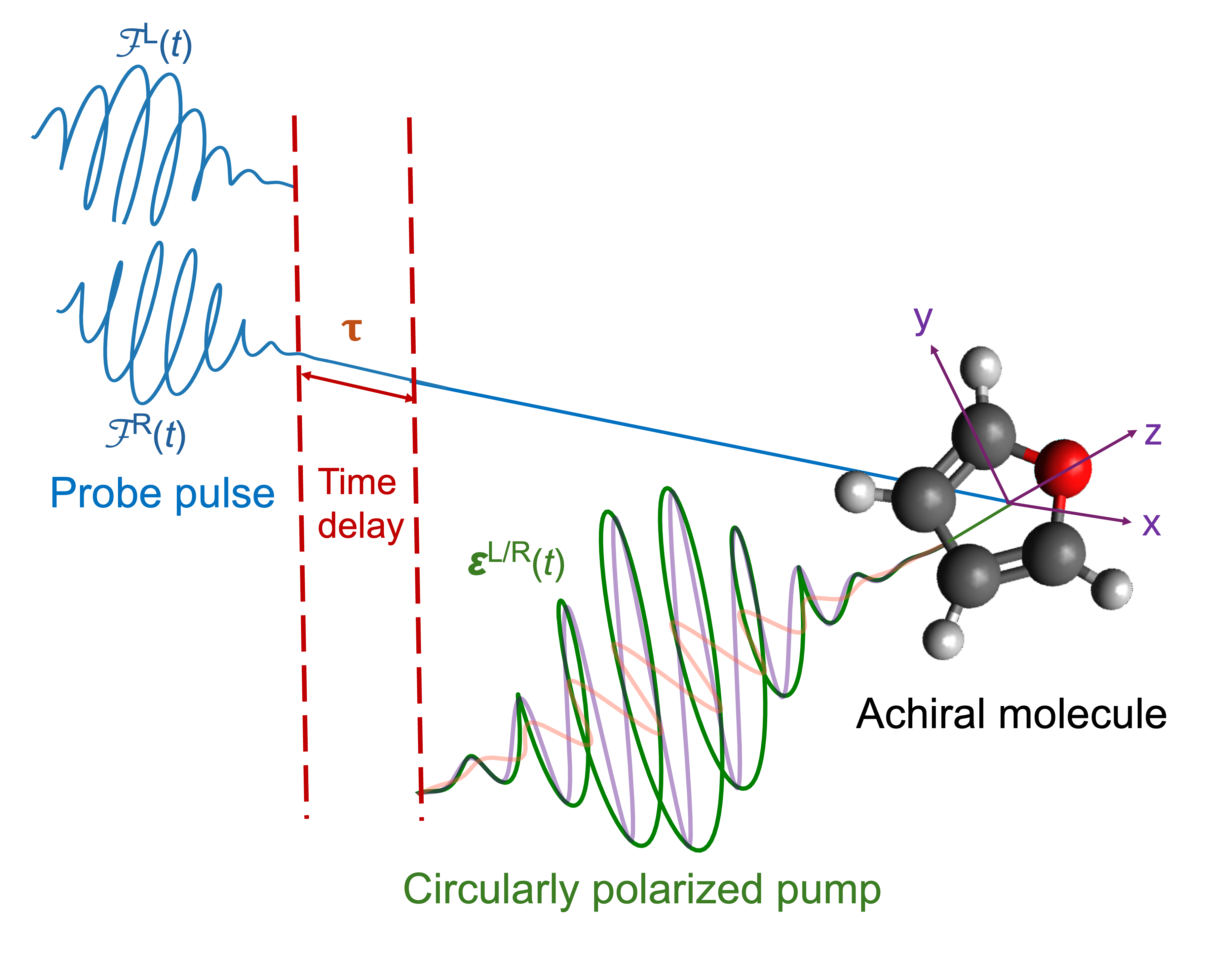}
    \caption{Schematic representation of pump-probe setup for TR-ECD spectral simulation, where a probe pulse $\bm{\mathcal{F}}^{\text{L/R}} (t)$ is applied at a time-delay of ($\tau$) after the end of pump pulse $ \bm{\mathcal{E}}^{\text{L/R}} (t) $.}
    \label{fig:pump-probe-setup}
\end{figure}
\begin{figure}[htb!]
	\centering
	\includegraphics[width=0.5\textwidth]{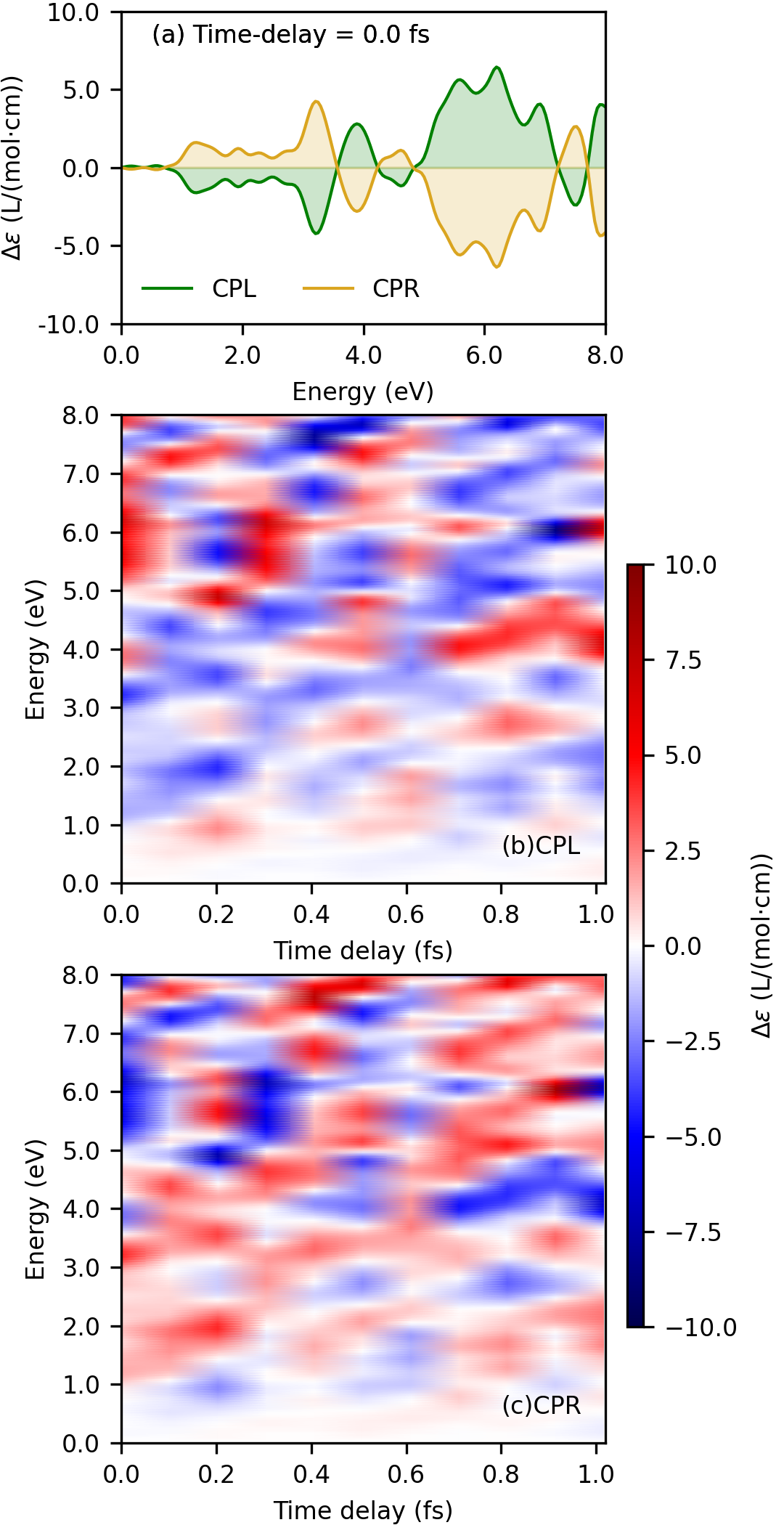}
	\caption{\fixme{Time-resolved electronic circular dichroism (TR-ECD) spectra shown as differential extinction coefficient ($\Delta \varepsilon$): (a) at zero pump–probe time delay ($\tau = 0$); and (b, c) as a function of time delay $\tau$ for (b) circularly polarized left (CPL) and (c) circularly polarized right (CPR) pump pulses.}
    }
	\label{fig:delays}
\end{figure}

Having established the generation of chiral persistent currents with attosecond characteristics, let us discuss their potential observation. A standard approach to spectroscopically detect the induced magnetic dipole moment in structurally chiral systems is through electronic circular dichroism (ECD) absorption spectroscopy, which measures the differential absorption of left and right circularly polarized light~\cite{2009_Barron,1980_ECD_Bouman_Hansen,2000_ecd_Berova,2012_WIRES_Warnke_Theory-ECD,2020_JCE_CD-concept}. We extend this technique to a typical pump-probe setup for obtaining time-resolved (TR-)ECD spectral signature; where the pump pulse, as discussed above, induces the electronic chirality in achiral oriented furan, while the probe pulse spectroscopically measures the induced chirality. This methodology forms the foundation of our proposed computational setup, as presented in Fig.~\ref{fig:pump-probe-setup}.

Theoretically, the ECD spectral function is calculated in the weak-probe regime from the imaginary part of the Rosenfeld tensor $\bm{\beta}(\omega)$~\cite{rosenfeld1929,condon1937,1980_ECD_Bouman_Hansen,2000_ecd_Berova,2009_Barron}. In the real-time TDDFT framework, this tensor can be obtained by the Fourier transformation of the time-dependent induced magnetic dipole moment  recorded during time propagation~\cite{2009_PCCP_Rubio_ECD,2016_JCP_XLi_ECD,2018_JCP_Repisky_RI-RT-chiroptical} 
\begin{align}
  \label{eq:5}
  \beta_{kj}(\omega)
  & =
  \frac{i}{\omega\mathcal{F}_{0}f_{k}}  
  \int_{-\infty}^{\infty}
  dt~m_j^{\text{ind}}(t) e^{i\omega t},
  \quad k,j\in x,y,z.
\end{align}
In this approach, the probe pulse is represented by an analytical Dirac delta function, $\bm{\mathcal{F}}(t) = \mathcal{F}_0 \bm{f} \delta (t-t_0)$, 
with amplitude $\mathcal{F}_0$, polarization vector $\bm{f}$, and initial time $t_0$. We have extended this formalism to TR-ECD with the delta probe pulse applied at $t_0=T+\tau$, where $T$ is the pump pulse duration and $\tau$ is the time delay between pump and probe pulses. To isolate the response of the chiral electronic wavepacket to the probe pulse only, we subtract the magnetic dipole moment induced by the pump pulse, in analogy with transient absorption spectroscopy~\cite{2023_JPCL_Repisky_RT-TAS}. The essential theoretical background is discussed in details in Supporting Information, Section S1.

The evolution of the chiral electronic wavepacket is monitored by varying the time-delay between pump and probe pulses, as shown in Fig.~\ref{fig:delays}.
TR-ECD spectra recorded immediately after the end of pump pulse displays mirror-image symmetry, as shown in Fig.~\ref{fig:delays}a, indicating an enantiomeric relationship between the CPL and CPR induced wavepackets.
Figures.~\ref{fig:delays}b and~\ref{fig:delays}c illustrate the
full TR-ECD spectra as obtained with CPL (left) and CPR (right) light, respectively. The spectra are rich in information and the primary observations are as follows:  (i) The mirror-image relationship between the CPL and CPR 2D signals is preserved over time. In other words, the helicity of the pump pulse induces an enantiomer-like relationship in the electronic wavepacket, and this relationship persists even during the rapid evolution of the chiral wavepacket. This observation is a direct consequence of the time-dependence of the induced magnetic moment shown in Fig.~\ref{fig:m_t}a, and the mirror relationship between CPR and CPL driving.  
(ii) The system exhibits chirality flips (sign reversal of TR-ECD spectral lines) generally on attosecond timescales observed on horizontal cuts in Figs.~\ref{fig:delays}b and~\ref{fig:delays}c. 
Notably, periods of these flips slightly vary for each energy window, likely connecting with various harmonics of the induced magnetic dipole moments presented in Fig.~\ref{fig:m_t}. Furthermore, we also investigated TR-ECD spectra of other oriented molecular systems, namely benzene and aniline. These results are \fixme{relegated} to the Supporting Information, Section S6, and show similar chiral imprints induced by CPL and CPR pump pulses. These findings suggest that light-induced attosecond persistent electronic chirality in achiral systems is a robust and transferable phenomenon.

In summary, our work explored the ultrafast chiral dynamics of oriented achiral molecules driven by a chiral pump pulse, demonstrating how electronic currents and their induced magnetic dipole moments encode a time-resolved chiral signature and persistent chiral molecular states. The induced magnetic dipole moment shows non-vanishing oscillations even after the end of the pulse, maintaining an enantiomer-like symmetry between CPL and CPR-induced electronic wavepackets. \fixme{Therefore, flips of induced magnetic moments and the resulting magnetic fields may arise from flips of chiral (this work) as well as from achiral~\cite{barth2010quantum} electronic currents induced by circularly polarized laser pulses.} 
An imprint of this oscillatory \fixme{behavior} of the current densities and magnetic dipole moments permits experimental observation by means of time-resolved electronic circular dichroism spectroscopy. We observe that the chiral response (both the magnetic dipole moment and the spectral function) does not oscillate with a characteristic frequency, but rather emerges from a hierarchy of harmonic frequencies coherent with the pump laser. Unlike conventional HHG from time-dependent electric dipole moment, the observed harmonics in the magnetic dipole moment stem from bound electronic currents, presenting an alternative route for probing and controlling chiral states, down to attosecond timescales. This method provides a spectral fingerprint for chiral wavepacket evolution, offering a powerful tool to study attosecond chiral dynamics in non-ionized regimes
\fixme{and without changing achiral nuclear configuration}.
Looking ahead, we believe that our results will motivate future studies in connected attosecond charge migration~\cite{2015_Science_Kraus_iodoacetylene,2010_Nature_RT-electron_motion,2014_Science_Calegari_phenylalanine,2022_PRR_solitons},
systems that are usually studied with photoemission and photo-dissociation, potentially linking to energy transfer mechanisms. The induced chiral transients could also be employed for plethora of technological applications, from high selectivity ultrafast control of chemical reactions, to light-induced phase transitions~\cite{2016_bisoyi_ChemRev}, electronics~\cite{2011_Morgen_chiralityflip}, and spintronics~\cite{2012_JPCL_Naaman_CISS,2020_Naaman_ACR_CISS,2022_AdvMat_CISS_review}

{\noindent\large{\textbf{Acknowledgements \\}}}
The authors acknowledge the support received from the Research Council of Norway through
a Centre of Excellence Grant (no. 262695), 
research grant (no. 315822), 
mobility grants (no. 301864 and no. 314814). 
T.M. acknowledges the support from the Marie
Sklodowska-Curie individual postdoctoral fellowship (grant no. 101152113). In
addition, M.R. acknowledges funding from the European Union’s Horizon 2020
research and innovation program under the Marie Sklodowska-Curie Grant
Agreement no. 945478 (SASPRO2), the Slovak Research and Development Agency
(grant no. APVV-22-0488), VEGA no. 1/0670/24, and the EU NextGenerationEU through the Recovery and Resilience Plan for Slovakia under the project No. 09I05-03-V02-00034.
In addition, this work is supported by the Cluster of Excellence 'CUI: Advanced Imaging of Matter' of the Deutsche Forschungsgemeinschaft (DFG) - EXC 2056 - project ID 390715994. We acknowledge support from the Max Planck-New York City Center for Non-Equilibrium Quantum Phenomena. The Flatiron Institute is a division of the Simons Foundation.
We thank Sigma2 - The National
Infrastructure for High Performance Computing and Data Storage in Norway, grant
no. NN14654K, and
EuroHPC regular access grant no. EU-25-8 for the computational resources. 
We thank Stanislav Komorovsky for fruitful discussions about visualisation.


{\noindent\large{\textbf{Supporting Information \\}}}
The Supporting Information available:
(1) Methodology
(2) Computational setup
(3) Ground state depopulation
(4) Analysis of induced electric dipole moment
(5) Time evolution of induced charge and current density (video)
(6) TR-ECD spectra of benzene and aniline.

{\noindent\large{\textbf{Author contributions \\}}}
All authors conceptualized the project. 
T.M., L.K., M.K. and M.R. designed the algorithm and performed all necessary code implementations. 
T.M. performed all electronic structure calculations, curated data, and wrote the first draft of manuscript. 
All authors contributed to finalize the manuscript.
\bibliography{tcd}

\providecommand{\latin}[1]{#1}
\makeatletter
\providecommand{\doi}
  {\begingroup\let\do\@makeother\dospecials
  \catcode`\{=1 \catcode`\}=2 \doi@aux}
\providecommand{\doi@aux}[1]{\endgroup\texttt{#1}}
\makeatother
\providecommand*\mcitethebibliography{\thebibliography}
\csname @ifundefined\endcsname{endmcitethebibliography}
  {\let\endmcitethebibliography\endthebibliography}{}
\begin{mcitethebibliography}{64}
\providecommand*\natexlab[1]{#1}
\providecommand*\mciteSetBstSublistMode[1]{}
\providecommand*\mciteSetBstMaxWidthForm[2]{}
\providecommand*\mciteBstWouldAddEndPuncttrue
  {\def\EndOfBibitem{\unskip.}}
\providecommand*\mciteBstWouldAddEndPunctfalse
  {\let\EndOfBibitem\relax}
\providecommand*\mciteSetBstMidEndSepPunct[3]{}
\providecommand*\mciteSetBstSublistLabelBeginEnd[3]{}
\providecommand*\EndOfBibitem{}
\mciteSetBstSublistMode{f}
\mciteSetBstMaxWidthForm{subitem}{(\alph{mcitesubitemcount})}
\mciteSetBstSublistLabelBeginEnd
  {\mcitemaxwidthsubitemform\space}
  {\relax}
  {\relax}

\bibitem[Cahn \latin{et~al.}(1966)Cahn, Ingold, and
  Prelog]{1966_AngweChem_Chirality}
Cahn,~R.~S.; Ingold,~C.; Prelog,~V. Specification of Molecular Chirality.
  \emph{Angew.~Chem.} \textbf{1966}, \emph{5}, 385--415\relax
\mciteBstWouldAddEndPuncttrue
\mciteSetBstMidEndSepPunct{\mcitedefaultmidpunct}
{\mcitedefaultendpunct}{\mcitedefaultseppunct}\relax
\EndOfBibitem
\bibitem[Wagni{\`e}re(2007)]{2007_book_chirality}
Wagni{\`e}re,~G.~H. \emph{On chirality and the universal asymmetry: reflections
  on image and mirror image}; John Wiley \& Sons, 2007\relax
\mciteBstWouldAddEndPuncttrue
\mciteSetBstMidEndSepPunct{\mcitedefaultmidpunct}
{\mcitedefaultendpunct}{\mcitedefaultseppunct}\relax
\EndOfBibitem
\bibitem[Changenet and Hache(2022)Changenet, and Hache]{2022_EPJST_psECD-rev}
Changenet,~P.; Hache,~F. Recent advances in the development of ultrafast
  electronic circular dichroism for probing the conformational dynamics of
  biomolecules in solution. \emph{Eur.~Phys.~J.:~Spec.~Top.} \textbf{2022},
  \emph{232}, 2117--2129\relax
\mciteBstWouldAddEndPuncttrue
\mciteSetBstMidEndSepPunct{\mcitedefaultmidpunct}
{\mcitedefaultendpunct}{\mcitedefaultseppunct}\relax
\EndOfBibitem
\bibitem[Ayuso \latin{et~al.}(2022)Ayuso, Ordonez, and
  Smirnova]{2022_PCCP_Smirnova_UltrafastChirality-rev}
Ayuso,~D.; Ordonez,~A.~F.; Smirnova,~O. Ultrafast chirality: the road to
  efficient chiral measurements. \emph{Phys.~Chem.~Chem.~Phys.} \textbf{2022},
  \emph{24}, 26962--26991\relax
\mciteBstWouldAddEndPuncttrue
\mciteSetBstMidEndSepPunct{\mcitedefaultmidpunct}
{\mcitedefaultendpunct}{\mcitedefaultseppunct}\relax
\EndOfBibitem
\bibitem[Oppermann \latin{et~al.}(2022)Oppermann, Zinna, Lacour, and
  Chergui]{2022_NatChem_Chergui_Fe-chiral-fsdynamics}
Oppermann,~M.; Zinna,~F.; Lacour,~J.; Chergui,~M. Chiral control of
  spin-crossover dynamics in Fe(II) complexes. \emph{Nat.~Chem.} \textbf{2022},
  \emph{14}, 739--745\relax
\mciteBstWouldAddEndPuncttrue
\mciteSetBstMidEndSepPunct{\mcitedefaultmidpunct}
{\mcitedefaultendpunct}{\mcitedefaultseppunct}\relax
\EndOfBibitem
\bibitem[Oppermann \latin{et~al.}(2019)Oppermann, Bauer, Rossi, Zinna, Helbing,
  Lacour, and Chergui]{2019_Optica_Chergui_TR-ECD}
Oppermann,~M.; Bauer,~B.; Rossi,~T.; Zinna,~F.; Helbing,~J.; Lacour,~J.;
  Chergui,~M. Ultrafast broadband circular dichroism in the deep ultraviolet.
  \emph{Optica} \textbf{2019}, \emph{6}, 56--60\relax
\mciteBstWouldAddEndPuncttrue
\mciteSetBstMidEndSepPunct{\mcitedefaultmidpunct}
{\mcitedefaultendpunct}{\mcitedefaultseppunct}\relax
\EndOfBibitem
\bibitem[Beaulieu \latin{et~al.}(2017)Beaulieu, Comby, Clergerie, Caillat,
  Descamps, Dudovich, Fabre, G{\'e}neaux, L{\'e}gar{\'e}, Petit, \latin{et~al.}
  others]{2017_Science_chiral-ionization-camphor}
Beaulieu,~S.; Comby,~A.; Clergerie,~A.; Caillat,~J.; Descamps,~D.;
  Dudovich,~N.; Fabre,~B.; G{\'e}neaux,~R.; L{\'e}gar{\'e},~F.; Petit,~S.
  \latin{et~al.}  Attosecond-resolved photoionization of chiral molecules.
  \emph{Science} \textbf{2017}, \emph{358}, 1288--1294\relax
\mciteBstWouldAddEndPuncttrue
\mciteSetBstMidEndSepPunct{\mcitedefaultmidpunct}
{\mcitedefaultendpunct}{\mcitedefaultseppunct}\relax
\EndOfBibitem
\bibitem[Tikhonov \latin{et~al.}(2022)Tikhonov, Blech, Leibscher, Greenman,
  Schnell, and Koch]{tikhonov2022pump}
Tikhonov,~D.~S.; Blech,~A.; Leibscher,~M.; Greenman,~L.; Schnell,~M.;
  Koch,~C.~P. Pump-probe spectroscopy of chiral vibrational dynamics.
  \emph{Sci. Adv.} \textbf{2022}, \emph{8}, eade0311\relax
\mciteBstWouldAddEndPuncttrue
\mciteSetBstMidEndSepPunct{\mcitedefaultmidpunct}
{\mcitedefaultendpunct}{\mcitedefaultseppunct}\relax
\EndOfBibitem
\bibitem[Leibscher \latin{et~al.}(2024)Leibscher, Pozzoli, Blech, Sigalotti,
  Boscain, and Koch]{leibscher2024quantum}
Leibscher,~M.; Pozzoli,~E.; Blech,~A.; Sigalotti,~M.; Boscain,~U.; Koch,~C.~P.
  Quantum control of rovibrational dynamics and application to light-induced
  molecular chirality. \emph{Phys.~Rev.~A} \textbf{2024}, \emph{109},
  012810\relax
\mciteBstWouldAddEndPuncttrue
\mciteSetBstMidEndSepPunct{\mcitedefaultmidpunct}
{\mcitedefaultendpunct}{\mcitedefaultseppunct}\relax
\EndOfBibitem
\bibitem[Wanie \latin{et~al.}(2024)Wanie, Bloch, M{\aa}nsson, Colaizzi,
  Ryabchuk, Saraswathula, Ordonez, Ayuso, Smirnova, Trabattoni, \latin{et~al.}
  others]{2024_Nat_Wanie_attochirality}
Wanie,~V.; Bloch,~E.; M{\aa}nsson,~E.~P.; Colaizzi,~L.; Ryabchuk,~S.;
  Saraswathula,~K.; Ordonez,~A.~F.; Ayuso,~D.; Smirnova,~O.; Trabattoni,~A.
  \latin{et~al.}  Capturing electron-driven chiral dynamics in UV-excited
  molecules. \emph{Nature} \textbf{2024}, \emph{630}, 109--115\relax
\mciteBstWouldAddEndPuncttrue
\mciteSetBstMidEndSepPunct{\mcitedefaultmidpunct}
{\mcitedefaultendpunct}{\mcitedefaultseppunct}\relax
\EndOfBibitem
\bibitem[Chen \latin{et~al.}(2024)Chen, Haase, Manz, Wang, and
  Yang]{2024_NatComm_Yonggang_attochirality}
Chen,~Y.; Haase,~D.; Manz,~J.; Wang,~H.; Yang,~Y. From chiral laser pulses to
  femto-and attosecond electronic chirality flips in achiral molecules.
  \emph{Nat.~Comm.} \textbf{2024}, \emph{15}, 565\relax
\mciteBstWouldAddEndPuncttrue
\mciteSetBstMidEndSepPunct{\mcitedefaultmidpunct}
{\mcitedefaultendpunct}{\mcitedefaultseppunct}\relax
\EndOfBibitem
\bibitem[Forbes and Andrews(2021)Forbes, and Andrews]{2021_Forbes_OAM}
Forbes,~K.~A.; Andrews,~D.~L. Orbital angular momentum of twisted light:
  chirality and optical activity. \emph{J.~Phys.~Photonics} \textbf{2021},
  \emph{3}, 022007\relax
\mciteBstWouldAddEndPuncttrue
\mciteSetBstMidEndSepPunct{\mcitedefaultmidpunct}
{\mcitedefaultendpunct}{\mcitedefaultseppunct}\relax
\EndOfBibitem
\bibitem[Barth \latin{et~al.}(2006)Barth, Manz, Shigeta, and
  Yagi]{2006_JACS_barth_ringcurrents}
Barth,~I.; Manz,~J.; Shigeta,~Y.; Yagi,~K. Unidirectional electronic ring
  current driven by a few cycle circularly polarized laser pulse: quantum model
  simulations for Mg- porphyrin. \emph{J.~Am.~Chem.~Soc.} \textbf{2006},
  \emph{128}, 7043--7049\relax
\mciteBstWouldAddEndPuncttrue
\mciteSetBstMidEndSepPunct{\mcitedefaultmidpunct}
{\mcitedefaultendpunct}{\mcitedefaultseppunct}\relax
\EndOfBibitem
\bibitem[Barth and Manz(2006)Barth, and Manz]{barth2006periodic}
Barth,~I.; Manz,~J. Periodic electron circulation induced by circularly
  polarized laser pulses: Quantum model simulations for Mg porphyrin.
  \emph{Angew.~Chem.} \textbf{2006}, \emph{45}, 2962--2965\relax
\mciteBstWouldAddEndPuncttrue
\mciteSetBstMidEndSepPunct{\mcitedefaultmidpunct}
{\mcitedefaultendpunct}{\mcitedefaultseppunct}\relax
\EndOfBibitem
\bibitem[Hermann \latin{et~al.}(2016)Hermann, Liu, Manz, Paulus, Perez-Torres,
  Pohl, and Tremblay]{hermann2016multidirectional}
Hermann,~G.; Liu,~C.; Manz,~J.; Paulus,~B.; Perez-Torres,~J.~F.; Pohl,~V.;
  Tremblay,~J.~C. Multidirectional angular electronic flux during adiabatic
  attosecond charge migration in excited benzene. \emph{J.~Phys.~Chem.~A}
  \textbf{2016}, \emph{120}, 5360--5369\relax
\mciteBstWouldAddEndPuncttrue
\mciteSetBstMidEndSepPunct{\mcitedefaultmidpunct}
{\mcitedefaultendpunct}{\mcitedefaultseppunct}\relax
\EndOfBibitem
\bibitem[Eckart \latin{et~al.}(2018)Eckart, Kunitski, Richter, Hartung, Rist,
  Trinter, Fehre, Schlott, Henrichs, Schmidt, \latin{et~al.}
  others]{2018_NatPhys_Eckart_ringcurrents}
Eckart,~S.; Kunitski,~M.; Richter,~M.; Hartung,~A.; Rist,~J.; Trinter,~F.;
  Fehre,~K.; Schlott,~N.; Henrichs,~K.; Schmidt,~L. P.~H. \latin{et~al.}
  Ultrafast preparation and detection of ring currents in single atoms.
  \emph{Nat.~Phys.} \textbf{2018}, \emph{14}, 701--704\relax
\mciteBstWouldAddEndPuncttrue
\mciteSetBstMidEndSepPunct{\mcitedefaultmidpunct}
{\mcitedefaultendpunct}{\mcitedefaultseppunct}\relax
\EndOfBibitem
\bibitem[Neufeld and Cohen(2019)Neufeld, and Cohen]{2019_PRL_Ofer_ringcurrents}
Neufeld,~O.; Cohen,~O. Background-free measurement of ring currents by
  symmetry-breaking high-harmonic spectroscopy. \emph{Phys.~Rev.~Lett.}
  \textbf{2019}, \emph{123}, 103202\relax
\mciteBstWouldAddEndPuncttrue
\mciteSetBstMidEndSepPunct{\mcitedefaultmidpunct}
{\mcitedefaultendpunct}{\mcitedefaultseppunct}\relax
\EndOfBibitem
\bibitem[de~Las~Heras \latin{et~al.}(2023)de~Las~Heras, Bonaf{\'e},
  Hern{\'a}ndez-Garc{\'\i}a, Rubio, and Neufeld]{2023_JPCL_Ofer_ringcurrents}
de~Las~Heras,~A.; Bonaf{\'e},~F.~P.; Hern{\'a}ndez-Garc{\'\i}a,~C.; Rubio,~A.;
  Neufeld,~O. Tunable Tesla-scale magnetic attosecond pulses through
  ring-current gating. \emph{J.~Phys.~Chem.~Lett.} \textbf{2023}, \emph{14},
  11160--11167\relax
\mciteBstWouldAddEndPuncttrue
\mciteSetBstMidEndSepPunct{\mcitedefaultmidpunct}
{\mcitedefaultendpunct}{\mcitedefaultseppunct}\relax
\EndOfBibitem
\bibitem[Pengel \latin{et~al.}(2017)Pengel, Kerbstadt, Johannmeyer, Englert,
  Bayer, and Wollenhaupt]{2017_PRL_Pengel_vortices}
Pengel,~D.; Kerbstadt,~S.; Johannmeyer,~D.; Englert,~L.; Bayer,~T.;
  Wollenhaupt,~M. Electron vortices in femtosecond multiphoton ionization.
  \emph{Phys.~Rev.~Lett.} \textbf{2017}, \emph{118}, 053003\relax
\mciteBstWouldAddEndPuncttrue
\mciteSetBstMidEndSepPunct{\mcitedefaultmidpunct}
{\mcitedefaultendpunct}{\mcitedefaultseppunct}\relax
\EndOfBibitem
\bibitem[Pengel \latin{et~al.}(2017)Pengel, Kerbstadt, Englert, Bayer, and
  Wollenhaupt]{2017_PRA_Pengel_3dvortices}
Pengel,~D.; Kerbstadt,~S.; Englert,~L.; Bayer,~T.; Wollenhaupt,~M. Control of
  three-dimensional electron vortices from femtosecond multiphoton ionization.
  \emph{Phys.~Rev.~A} \textbf{2017}, \emph{96}, 043426\relax
\mciteBstWouldAddEndPuncttrue
\mciteSetBstMidEndSepPunct{\mcitedefaultmidpunct}
{\mcitedefaultendpunct}{\mcitedefaultseppunct}\relax
\EndOfBibitem
\bibitem[Ayuso \latin{et~al.}(2019)Ayuso, Neufeld, Ordonez, Decleva, Lerner,
  Cohen, Ivanov, and Smirnova]{2019_NatPhotonics_Smirnova_chirality}
Ayuso,~D.; Neufeld,~O.; Ordonez,~A.~F.; Decleva,~P.; Lerner,~G.; Cohen,~O.;
  Ivanov,~M.; Smirnova,~O. Synthetic chiral light for efficient control of
  chiral light--matter interaction. \emph{Nat.~Photonics.} \textbf{2019},
  \emph{13}, 866--871\relax
\mciteBstWouldAddEndPuncttrue
\mciteSetBstMidEndSepPunct{\mcitedefaultmidpunct}
{\mcitedefaultendpunct}{\mcitedefaultseppunct}\relax
\EndOfBibitem
\bibitem[Mayer \latin{et~al.}(2022)Mayer, Patchkovskii, Morales, Ivanov, and
  Smirnova]{2022_PRL_Smirnova_chirality}
Mayer,~N.; Patchkovskii,~S.; Morales,~F.; Ivanov,~M.; Smirnova,~O. Imprinting
  chirality on atoms using synthetic chiral light fields.
  \emph{Phys.~Rev.~Lett.} \textbf{2022}, \emph{129}, 243201\relax
\mciteBstWouldAddEndPuncttrue
\mciteSetBstMidEndSepPunct{\mcitedefaultmidpunct}
{\mcitedefaultendpunct}{\mcitedefaultseppunct}\relax
\EndOfBibitem
\bibitem[Habibovi{\'c} \latin{et~al.}(2024)Habibovi{\'c}, Hamilton, Neufeld,
  and Rego]{2024_NatRevPhys_Ofer_review}
Habibovi{\'c},~D.; Hamilton,~K.~R.; Neufeld,~O.; Rego,~L. Emerging tailored
  light sources for studying chirality and symmetry. \emph{Nat.~Rev.~Phys.}
  \textbf{2024}, \emph{6}, 663--675\relax
\mciteBstWouldAddEndPuncttrue
\mciteSetBstMidEndSepPunct{\mcitedefaultmidpunct}
{\mcitedefaultendpunct}{\mcitedefaultseppunct}\relax
\EndOfBibitem
\bibitem[Neufeld \latin{et~al.}(2019)Neufeld, Podolsky, and
  Cohen]{2019_Ofer_floquetHHG}
Neufeld,~O.; Podolsky,~D.; Cohen,~O. Floquet group theory and its application
  to selection rules in harmonic generation. \emph{Nat.~Comm.} \textbf{2019},
  \emph{10}, 405\relax
\mciteBstWouldAddEndPuncttrue
\mciteSetBstMidEndSepPunct{\mcitedefaultmidpunct}
{\mcitedefaultendpunct}{\mcitedefaultseppunct}\relax
\EndOfBibitem
\bibitem[Siegrist \latin{et~al.}(2019)Siegrist, Gessner, Ossiander, Denker,
  Chang, Schr{\"o}der, Guggenmos, Cui, Walowski, Martens, \latin{et~al.}
  others]{2019_Nature_Siegrist_magnetism}
Siegrist,~F.; Gessner,~J.~A.; Ossiander,~M.; Denker,~C.; Chang,~Y.-P.;
  Schr{\"o}der,~M.~C.; Guggenmos,~A.; Cui,~Y.; Walowski,~J.; Martens,~U.
  \latin{et~al.}  Light-wave dynamic control of magnetism. \emph{Nature}
  \textbf{2019}, \emph{571}, 240--244\relax
\mciteBstWouldAddEndPuncttrue
\mciteSetBstMidEndSepPunct{\mcitedefaultmidpunct}
{\mcitedefaultendpunct}{\mcitedefaultseppunct}\relax
\EndOfBibitem
\bibitem[Neufeld \latin{et~al.}(2023)Neufeld, Tancogne-Dejean, De~Giovannini,
  H{\"u}bener, and Rubio]{2023_npjComputationlMat_Ofer_magnetisation}
Neufeld,~O.; Tancogne-Dejean,~N.; De~Giovannini,~U.; H{\"u}bener,~H.; Rubio,~A.
  Attosecond magnetization dynamics in non-magnetic materials driven by intense
  femtosecond lasers. \emph{Npj~Comput.~Mater.} \textbf{2023}, \emph{9},
  39\relax
\mciteBstWouldAddEndPuncttrue
\mciteSetBstMidEndSepPunct{\mcitedefaultmidpunct}
{\mcitedefaultendpunct}{\mcitedefaultseppunct}\relax
\EndOfBibitem
\bibitem[Morgenstern(2011)]{2011_Morgen_chiralityflip}
Morgenstern,~K. Switching individual molecules by light and electrons: From
  isomerisation to chirality flip. \emph{Prog~Surf~Sci.} \textbf{2011},
  \emph{86}, 115--161\relax
\mciteBstWouldAddEndPuncttrue
\mciteSetBstMidEndSepPunct{\mcitedefaultmidpunct}
{\mcitedefaultendpunct}{\mcitedefaultseppunct}\relax
\EndOfBibitem
\bibitem[Evers \latin{et~al.}(2022)Evers, Aharony, Bar-Gill, Entin-Wohlman,
  Hedeg{\aa}rd, Hod, Jelinek, Kamieniarz, Lemeshko, Michaeli, \latin{et~al.}
  others]{2022_AdvMat_CISS_review}
Evers,~F.; Aharony,~A.; Bar-Gill,~N.; Entin-Wohlman,~O.; Hedeg{\aa}rd,~P.;
  Hod,~O.; Jelinek,~P.; Kamieniarz,~G.; Lemeshko,~M.; Michaeli,~K.
  \latin{et~al.}  Theory of chirality induced spin selectivity: Progress and
  challenges. \emph{Adv.~Mater.} \textbf{2022}, \emph{34}, 2106629\relax
\mciteBstWouldAddEndPuncttrue
\mciteSetBstMidEndSepPunct{\mcitedefaultmidpunct}
{\mcitedefaultendpunct}{\mcitedefaultseppunct}\relax
\EndOfBibitem
\bibitem[Naaman and Waldeck(2012)Naaman, and Waldeck]{2012_JPCL_Naaman_CISS}
Naaman,~R.; Waldeck,~D.~H. Chiral-Induced Spin Selectivity Effect.
  \emph{J.~Phys.~Chem.~Lett.} \textbf{2012}, \emph{3}, 2178--2187\relax
\mciteBstWouldAddEndPuncttrue
\mciteSetBstMidEndSepPunct{\mcitedefaultmidpunct}
{\mcitedefaultendpunct}{\mcitedefaultseppunct}\relax
\EndOfBibitem
\bibitem[Bisoyi and Li(2016)Bisoyi, and Li]{2016_bisoyi_ChemRev}
Bisoyi,~H.~K.; Li,~Q. Light-driven liquid crystalline materials: from
  photo-induced phase transitions and property modulations to applications.
  \emph{Chem.~Rev.} \textbf{2016}, \emph{116}, 15089--15166\relax
\mciteBstWouldAddEndPuncttrue
\mciteSetBstMidEndSepPunct{\mcitedefaultmidpunct}
{\mcitedefaultendpunct}{\mcitedefaultseppunct}\relax
\EndOfBibitem
\bibitem[Fleischer \latin{et~al.}(2011)Fleischer, Zhou, Field, and
  Nelson]{2011_PRL_Fleischer_orientation}
Fleischer,~S.; Zhou,~Y.; Field,~R.~W.; Nelson,~K.~A. Molecular orientation and
  alignment by intense single-cycle THz pulses. \emph{Phys.~Rev.~Lett.}
  \textbf{2011}, \emph{107}, 163603\relax
\mciteBstWouldAddEndPuncttrue
\mciteSetBstMidEndSepPunct{\mcitedefaultmidpunct}
{\mcitedefaultendpunct}{\mcitedefaultseppunct}\relax
\EndOfBibitem
\bibitem[Fuji \latin{et~al.}(2010)Fuji, Suzuki, Horio, Suzuki, Mitri{\'c},
  Werner, and Bona{\v{c}}i{\'c}-Kouteck{\`y}]{2010_JCP_Fuji_furan}
Fuji,~T.; Suzuki,~Y.-I.; Horio,~T.; Suzuki,~T.; Mitri{\'c},~R.; Werner,~U.;
  Bona{\v{c}}i{\'c}-Kouteck{\`y},~V. Ultrafast photodynamics of furan.
  \emph{J.~Chem.~Phys.} \textbf{2010}, \emph{133}, 234303\relax
\mciteBstWouldAddEndPuncttrue
\mciteSetBstMidEndSepPunct{\mcitedefaultmidpunct}
{\mcitedefaultendpunct}{\mcitedefaultseppunct}\relax
\EndOfBibitem
\bibitem[Liu \latin{et~al.}(2015)Liu, Knopp, Qin, and
  Gerber]{2015_CP_Liu_furan}
Liu,~Y.; Knopp,~G.; Qin,~C.; Gerber,~T. Tracking ultrafast relaxation dynamics
  of furan by femtosecond photoelectron imaging. \emph{Chem.~Phys.}
  \textbf{2015}, \emph{446}, 142--147\relax
\mciteBstWouldAddEndPuncttrue
\mciteSetBstMidEndSepPunct{\mcitedefaultmidpunct}
{\mcitedefaultendpunct}{\mcitedefaultseppunct}\relax
\EndOfBibitem
\bibitem[Hua \latin{et~al.}(2016)Hua, Oesterling, Biggs, Zhang, Ando,
  de~Vivie-Riedle, Fingerhut, and Mukamel]{2016_SD_Hua_furan}
Hua,~W.; Oesterling,~S.; Biggs,~J.~D.; Zhang,~Y.; Ando,~H.;
  de~Vivie-Riedle,~R.; Fingerhut,~B.~P.; Mukamel,~S. Monitoring conical
  intersections in the ring opening of furan by attosecond stimulated X-ray
  Raman spectroscopy. \emph{Struct.~Dyn.} \textbf{2016}, \emph{3}, 023601\relax
\mciteBstWouldAddEndPuncttrue
\mciteSetBstMidEndSepPunct{\mcitedefaultmidpunct}
{\mcitedefaultendpunct}{\mcitedefaultseppunct}\relax
\EndOfBibitem
\bibitem[Sun \latin{et~al.}(2024)Sun, Gu, Hu, Lu, Tang, Chernyak, Li, and
  Mukamel]{2024_JACS_Mukamel_TRMCD}
Sun,~S.; Gu,~B.; Hu,~H.; Lu,~L.; Tang,~D.; Chernyak,~V.~Y.; Li,~X.; Mukamel,~S.
  Direct Probe of Conical Intersection Photochemistry by Time-Resolved X-ray
  Magnetic Circular Dichroism. \emph{J.~Am.~Chem.~Soc.} \textbf{2024},
  \emph{146}, 19863–19873\relax
\mciteBstWouldAddEndPuncttrue
\mciteSetBstMidEndSepPunct{\mcitedefaultmidpunct}
{\mcitedefaultendpunct}{\mcitedefaultseppunct}\relax
\EndOfBibitem
\bibitem[Uenishi \latin{et~al.}(2024)Uenishi, Boyer, Karashima, Humeniuk, and
  Suzuki]{2024_JPCL_Uenishi_furan}
Uenishi,~R.; Boyer,~A.; Karashima,~S.; Humeniuk,~A.; Suzuki,~T. Signatures of
  Conical Intersections in Extreme Ultraviolet Photoelectron Spectra of Furan
  Measured with 15 fs Time Resolution. \emph{J.~Phys.~Chem.~Lett.}
  \textbf{2024}, \emph{15}, 2222--2227\relax
\mciteBstWouldAddEndPuncttrue
\mciteSetBstMidEndSepPunct{\mcitedefaultmidpunct}
{\mcitedefaultendpunct}{\mcitedefaultseppunct}\relax
\EndOfBibitem
\bibitem[Repisky \latin{et~al.}(2020)Repisky, Komorovsky, Kadek, Konecny,
  Ekstr{\"o}m, Malkin, Kaupp, Ruud, Malkina, and Malkin]{2020_JCP_ReSpect}
Repisky,~M.; Komorovsky,~S.; Kadek,~M.; Konecny,~L.; Ekstr{\"o}m,~U.;
  Malkin,~E.; Kaupp,~M.; Ruud,~K.; Malkina,~O.~L.; Malkin,~V.~G. ReSpect:
  Relativistic spectroscopy DFT program package. \emph{J.~Chem.~Phys.}
  \textbf{2020}, \emph{152}, 184101\relax
\mciteBstWouldAddEndPuncttrue
\mciteSetBstMidEndSepPunct{\mcitedefaultmidpunct}
{\mcitedefaultendpunct}{\mcitedefaultseppunct}\relax
\EndOfBibitem
\bibitem[Repisky \latin{et~al.}(2025)Repisky, Komorovsky, Konecny, Kadek,
  Moitra, Joosten, Misenkova, Vikhamar-Sandberg, Kaupp, Ruud, Malkina, and
  Malkin]{repisky2025respect}
Repisky,~M.; Komorovsky,~S.; Konecny,~L.; Kadek,~M.; Moitra,~T.; Joosten,~M.;
  Misenkova,~D.; Vikhamar-Sandberg,~R.; Kaupp,~M.; Ruud,~K. \latin{et~al.}  X2C
  Hamiltonian Models in ReSpect: Bridging Accuracy and Efficiency. \emph{arXiv
  preprint arXiv:2505.01088} \textbf{2025}, \relax
\mciteBstWouldAddEndPunctfalse
\mciteSetBstMidEndSepPunct{\mcitedefaultmidpunct}
{}{\mcitedefaultseppunct}\relax
\EndOfBibitem
\bibitem[Dunning~Jr(1989)]{1989_dunning_basis_1}
Dunning~Jr,~T.~H. Gaussian basis sets for use in correlated molecular
  calculations. I. The atoms boron through neon and hydrogen.
  \emph{J.~Chem.~Phys.} \textbf{1989}, \emph{90}, 1007--1023\relax
\mciteBstWouldAddEndPuncttrue
\mciteSetBstMidEndSepPunct{\mcitedefaultmidpunct}
{\mcitedefaultendpunct}{\mcitedefaultseppunct}\relax
\EndOfBibitem
\bibitem[Kendall \latin{et~al.}(1992)Kendall, Dunning, and
  Harrison]{1992_dunning_basis_2}
Kendall,~R.~A.; Dunning,~T.~H.; Harrison,~R.~J. Electron affinities of the
  first-row atoms revisited. Systematic basis sets and wave functions.
  \emph{J.~Chem.~Phys.} \textbf{1992}, \emph{96}, 6796--6806\relax
\mciteBstWouldAddEndPuncttrue
\mciteSetBstMidEndSepPunct{\mcitedefaultmidpunct}
{\mcitedefaultendpunct}{\mcitedefaultseppunct}\relax
\EndOfBibitem
\bibitem[Adamo and Barone(1999)Adamo, and Barone]{PBE0}
Adamo,~C.; Barone,~V. {Toward reliable density functional methods without
  adjustable parameters: The PBE0 model}. \emph{J.~Chem.~Phys.} \textbf{1999},
  \emph{110}, 6158--6170\relax
\mciteBstWouldAddEndPuncttrue
\mciteSetBstMidEndSepPunct{\mcitedefaultmidpunct}
{\mcitedefaultendpunct}{\mcitedefaultseppunct}\relax
\EndOfBibitem
\bibitem[Liu \latin{et~al.}(2024)Liu, Manz, Wang, and Yang]{liu2024cpc}
Liu,~G.; Manz,~J.; Wang,~H.; Yang,~Y. Spatio-Temporal Symmetries of Electronic
  Chirality Flips in Oriented RbCs Induced by two Coincident Laser Pulses with
  Circular++,+-,-+,-- Polarizations. \emph{ChemPhysChem.} \textbf{2024},
  \emph{25}, e202400595\relax
\mciteBstWouldAddEndPuncttrue
\mciteSetBstMidEndSepPunct{\mcitedefaultmidpunct}
{\mcitedefaultendpunct}{\mcitedefaultseppunct}\relax
\EndOfBibitem
\bibitem[Cireasa \latin{et~al.}(2015)Cireasa, Boguslavskiy, Pons, Wong,
  Descamps, Petit, Ruf, Thir{\'e}, Ferr{\'e}, Suarez, \latin{et~al.}
  others]{2015_Bharadwaj_NatPhys_cHHG}
Cireasa,~R.; Boguslavskiy,~A.; Pons,~B.; Wong,~M.; Descamps,~D.; Petit,~S.;
  Ruf,~H.; Thir{\'e},~N.; Ferr{\'e},~A.; Suarez,~J. \latin{et~al.}  Probing
  molecular chirality on a sub-femtosecond timescale. \emph{Nat.~Phys.}
  \textbf{2015}, \emph{11}, 654--658\relax
\mciteBstWouldAddEndPuncttrue
\mciteSetBstMidEndSepPunct{\mcitedefaultmidpunct}
{\mcitedefaultendpunct}{\mcitedefaultseppunct}\relax
\EndOfBibitem
\bibitem[Neufeld \latin{et~al.}(2019)Neufeld, Ayuso, Decleva, Ivanov, Smirnova,
  and Cohen]{2019_Ofer_PRX_HHG_chiral}
Neufeld,~O.; Ayuso,~D.; Decleva,~P.; Ivanov,~M.~Y.; Smirnova,~O.; Cohen,~O.
  Ultrasensitive Chiral Spectroscopy by Dynamical Symmetry Breaking in High
  Harmonic Generation. \emph{Phys.~Rev.~X.} \textbf{2019}, \emph{9},
  031002\relax
\mciteBstWouldAddEndPuncttrue
\mciteSetBstMidEndSepPunct{\mcitedefaultmidpunct}
{\mcitedefaultendpunct}{\mcitedefaultseppunct}\relax
\EndOfBibitem
\bibitem[Ayuso \latin{et~al.}(2022)Ayuso, Ordonez, Decleva, Ivanov, and
  Smirnova]{2022_Ayuso_Chiral_HHG}
Ayuso,~D.; Ordonez,~A.~F.; Decleva,~P.; Ivanov,~M.; Smirnova,~O. Strong chiral
  response in non-collinear high harmonic generation driven by purely
  electric-dipole interactions. \emph{Opt.~Express} \textbf{2022}, \emph{30},
  4659--4667\relax
\mciteBstWouldAddEndPuncttrue
\mciteSetBstMidEndSepPunct{\mcitedefaultmidpunct}
{\mcitedefaultendpunct}{\mcitedefaultseppunct}\relax
\EndOfBibitem
\bibitem[Huang \latin{et~al.}(2018)Huang, Hern{\'a}ndez-Garc{\'\i}a, Huang,
  Huang, Lu, Rego, Hickstein, Ellis, Jaron-Becker, Becker, \latin{et~al.}
  others]{2018_NatPhotonics_polarization}
Huang,~P.-C.; Hern{\'a}ndez-Garc{\'\i}a,~C.; Huang,~J.-T.; Huang,~P.-Y.;
  Lu,~C.-H.; Rego,~L.; Hickstein,~D.~D.; Ellis,~J.~L.; Jaron-Becker,~A.;
  Becker,~A. \latin{et~al.}  Polarization control of isolated high-harmonic
  pulses. \emph{Nat.~Photonics.} \textbf{2018}, \emph{12}, 349--354\relax
\mciteBstWouldAddEndPuncttrue
\mciteSetBstMidEndSepPunct{\mcitedefaultmidpunct}
{\mcitedefaultendpunct}{\mcitedefaultseppunct}\relax
\EndOfBibitem
\bibitem[Barron(2009)]{2009_Barron}
Barron,~L.~D. \emph{Molecular light scattering and optical activity}; Cambridge
  University Press, 2009\relax
\mciteBstWouldAddEndPuncttrue
\mciteSetBstMidEndSepPunct{\mcitedefaultmidpunct}
{\mcitedefaultendpunct}{\mcitedefaultseppunct}\relax
\EndOfBibitem
\bibitem[Hansen and Bouman(1980)Hansen, and Bouman]{1980_ECD_Bouman_Hansen}
Hansen,~A.~E.; Bouman,~T.~D. Natural chiroptical spectroscopy: theory and
  computations. \emph{Adv.~Chem.~Phys.} \textbf{1980}, 545--644\relax
\mciteBstWouldAddEndPuncttrue
\mciteSetBstMidEndSepPunct{\mcitedefaultmidpunct}
{\mcitedefaultendpunct}{\mcitedefaultseppunct}\relax
\EndOfBibitem
\bibitem[Berova \latin{et~al.}(2000)Berova, Nakanishi, and
  Woody]{2000_ecd_Berova}
Berova,~N.; Nakanishi,~K.; Woody,~R.~W. \emph{Circular dichroism: principles
  and applications}; John Wiley \& Sons, 2000\relax
\mciteBstWouldAddEndPuncttrue
\mciteSetBstMidEndSepPunct{\mcitedefaultmidpunct}
{\mcitedefaultendpunct}{\mcitedefaultseppunct}\relax
\EndOfBibitem
\bibitem[Warnke and Furche(2012)Warnke, and
  Furche]{2012_WIRES_Warnke_Theory-ECD}
Warnke,~I.; Furche,~F. Circular dichroism: electronic.
  \emph{Wiley~Interdiscip.~Rev.~Comput.~Mol.~Sci.} \textbf{2012}, \emph{2},
  150--166\relax
\mciteBstWouldAddEndPuncttrue
\mciteSetBstMidEndSepPunct{\mcitedefaultmidpunct}
{\mcitedefaultendpunct}{\mcitedefaultseppunct}\relax
\EndOfBibitem
\bibitem[Andrews and Tretton(2020)Andrews, and Tretton]{2020_JCE_CD-concept}
Andrews,~S.~S.; Tretton,~J. Physical principles of circular dichroism.
  \emph{J.~Chem.~Edu.} \textbf{2020}, \emph{97}, 4370--4376\relax
\mciteBstWouldAddEndPuncttrue
\mciteSetBstMidEndSepPunct{\mcitedefaultmidpunct}
{\mcitedefaultendpunct}{\mcitedefaultseppunct}\relax
\EndOfBibitem
\bibitem[Rosenfeld(1929)]{rosenfeld1929}
Rosenfeld,~L. Quantenmechanische Theorie der nat{\"u}rlichen optischen
  Aktivit{\"a}t von Fl{\"u}ssigkeiten und Gasen. \emph{Zeitschrift f{\"u}r
  Physik} \textbf{1929}, \emph{52}, 161--174\relax
\mciteBstWouldAddEndPuncttrue
\mciteSetBstMidEndSepPunct{\mcitedefaultmidpunct}
{\mcitedefaultendpunct}{\mcitedefaultseppunct}\relax
\EndOfBibitem
\bibitem[Condon(1937)]{condon1937}
Condon,~E.~U. Theories of optical rotatory power. \emph{Rev.~Mod.~Phys.}
  \textbf{1937}, \emph{9}, 432\relax
\mciteBstWouldAddEndPuncttrue
\mciteSetBstMidEndSepPunct{\mcitedefaultmidpunct}
{\mcitedefaultendpunct}{\mcitedefaultseppunct}\relax
\EndOfBibitem
\bibitem[Varsano \latin{et~al.}(2009)Varsano, Espinosa-Leal, Andrade, Marques,
  Di~Felice, and Rubio]{2009_PCCP_Rubio_ECD}
Varsano,~D.; Espinosa-Leal,~L.~A.; Andrade,~X.; Marques,~M.~A.; Di~Felice,~R.;
  Rubio,~A. Towards a gauge invariant method for molecular chiroptical
  properties in TDDFT. \emph{Phys.~Chem.~Chem.~Phys.} \textbf{2009}, \emph{11},
  4481--4489\relax
\mciteBstWouldAddEndPuncttrue
\mciteSetBstMidEndSepPunct{\mcitedefaultmidpunct}
{\mcitedefaultendpunct}{\mcitedefaultseppunct}\relax
\EndOfBibitem
\bibitem[Goings and Li(2016)Goings, and Li]{2016_JCP_XLi_ECD}
Goings,~J.~J.; Li,~X. An atomic orbital based real-time time-dependent density
  functional theory for computing electronic circular dichroism band spectra.
  \emph{J.~Chem.~Phys.} \textbf{2016}, \emph{144}, 234102\relax
\mciteBstWouldAddEndPuncttrue
\mciteSetBstMidEndSepPunct{\mcitedefaultmidpunct}
{\mcitedefaultendpunct}{\mcitedefaultseppunct}\relax
\EndOfBibitem
\bibitem[Konecny \latin{et~al.}(2018)Konecny, Kadek, Komorovsky, Ruud, and
  Repisky]{2018_JCP_Repisky_RI-RT-chiroptical}
Konecny,~L.; Kadek,~M.; Komorovsky,~S.; Ruud,~K.; Repisky,~M.
  Resolution-of-identity accelerated relativistic two-and four-component
  electron dynamics approach to chiroptical spectroscopies.
  \emph{J.~Chem.~Phys.} \textbf{2018}, \emph{149}, 204104\relax
\mciteBstWouldAddEndPuncttrue
\mciteSetBstMidEndSepPunct{\mcitedefaultmidpunct}
{\mcitedefaultendpunct}{\mcitedefaultseppunct}\relax
\EndOfBibitem
\bibitem[Moitra \latin{et~al.}(2023)Moitra, Konecny, Kadek, Rubio, and
  Repisky]{2023_JPCL_Repisky_RT-TAS}
Moitra,~T.; Konecny,~L.; Kadek,~M.; Rubio,~A.; Repisky,~M. Accurate
  Relativistic Real-Time Time-Dependent Density Functional Theory for Valence
  and Core Attosecond Transient Absorption Spectroscopy.
  \emph{J.~Phys.~Chem.~Lett.} \textbf{2023}, \emph{14}, 1714--1724\relax
\mciteBstWouldAddEndPuncttrue
\mciteSetBstMidEndSepPunct{\mcitedefaultmidpunct}
{\mcitedefaultendpunct}{\mcitedefaultseppunct}\relax
\EndOfBibitem
\bibitem[Barth and Manz(2010)Barth, and Manz]{barth2010quantum}
Barth,~I.; Manz,~J. \emph{Progress in Ultrafast Intense Laser Science VI};
  Springer, 2010; pp 21--44\relax
\mciteBstWouldAddEndPuncttrue
\mciteSetBstMidEndSepPunct{\mcitedefaultmidpunct}
{\mcitedefaultendpunct}{\mcitedefaultseppunct}\relax
\EndOfBibitem
\bibitem[Kraus \latin{et~al.}(2015)Kraus, Mignolet, Baykusheva, Rupenyan,
  Horn{\`y}, Penka, Grassi, Tolstikhin, Schneider, Jensen, \latin{et~al.}
  others]{2015_Science_Kraus_iodoacetylene}
Kraus,~P.~M.; Mignolet,~B.; Baykusheva,~D.; Rupenyan,~A.; Horn{\`y},~L.;
  Penka,~E.~F.; Grassi,~G.; Tolstikhin,~O.~I.; Schneider,~J.; Jensen,~F.
  \latin{et~al.}  Measurement and laser control of attosecond charge migration
  in ionized iodoacetylene. \emph{Science} \textbf{2015}, \emph{350},
  790--795\relax
\mciteBstWouldAddEndPuncttrue
\mciteSetBstMidEndSepPunct{\mcitedefaultmidpunct}
{\mcitedefaultendpunct}{\mcitedefaultseppunct}\relax
\EndOfBibitem
\bibitem[Goulielmakis \latin{et~al.}(2010)Goulielmakis, Loh, Wirth, Santra,
  Rohringer, Yakovlev, Zherebtsov, Pfeifer, Azzeer, Kling, \latin{et~al.}
  others]{2010_Nature_RT-electron_motion}
Goulielmakis,~E.; Loh,~Z.-H.; Wirth,~A.; Santra,~R.; Rohringer,~N.;
  Yakovlev,~V.~S.; Zherebtsov,~S.; Pfeifer,~T.; Azzeer,~A.~M.; Kling,~M.~F.
  \latin{et~al.}  Real-time observation of valence electron motion.
  \emph{Nature} \textbf{2010}, \emph{466}, 739--743\relax
\mciteBstWouldAddEndPuncttrue
\mciteSetBstMidEndSepPunct{\mcitedefaultmidpunct}
{\mcitedefaultendpunct}{\mcitedefaultseppunct}\relax
\EndOfBibitem
\bibitem[Calegari \latin{et~al.}(2014)Calegari, Ayuso, Trabattoni, Belshaw,
  De~Camillis, Anumula, Frassetto, Poletto, Palacios, Decleva, \latin{et~al.}
  others]{2014_Science_Calegari_phenylalanine}
Calegari,~F.; Ayuso,~D.; Trabattoni,~A.; Belshaw,~L.; De~Camillis,~S.;
  Anumula,~S.; Frassetto,~F.; Poletto,~L.; Palacios,~A.; Decleva,~P.
  \latin{et~al.}  Ultrafast electron dynamics in phenylalanine initiated by
  attosecond pulses. \emph{Science} \textbf{2014}, \emph{346}, 336--339\relax
\mciteBstWouldAddEndPuncttrue
\mciteSetBstMidEndSepPunct{\mcitedefaultmidpunct}
{\mcitedefaultendpunct}{\mcitedefaultseppunct}\relax
\EndOfBibitem
\bibitem[Mauger \latin{et~al.}(2022)Mauger, Folorunso, Hamer, Chandre, Gaarde,
  Lopata, and Schafer]{2022_PRR_solitons}
Mauger,~F.; Folorunso,~A.~S.; Hamer,~K.~A.; Chandre,~C.; Gaarde,~M.~B.;
  Lopata,~K.; Schafer,~K.~J. Charge migration and attosecond solitons in
  conjugated organic molecules. \emph{Phys.~Rev.~Res.} \textbf{2022}, \emph{4},
  013073\relax
\mciteBstWouldAddEndPuncttrue
\mciteSetBstMidEndSepPunct{\mcitedefaultmidpunct}
{\mcitedefaultendpunct}{\mcitedefaultseppunct}\relax
\EndOfBibitem
\bibitem[Naaman \latin{et~al.}(2020)Naaman, Paltiel, and
  Waldeck]{2020_Naaman_ACR_CISS}
Naaman,~R.; Paltiel,~Y.; Waldeck,~D.~H. Chiral induced spin selectivity gives a
  new twist on spin-control in chemistry. \emph{Acc.~Chem.~Res.} \textbf{2020},
  \emph{53}, 2659--2667\relax
\mciteBstWouldAddEndPuncttrue
\mciteSetBstMidEndSepPunct{\mcitedefaultmidpunct}
{\mcitedefaultendpunct}{\mcitedefaultseppunct}\relax
\EndOfBibitem
\end{mcitethebibliography}

\clearpage
\newgeometry{top=2cm, bottom=2cm, left=2cm, right=2cm}

\setcounter{figure}{0}
\setcounter{section}{0}
\setcounter{table}{0}
\setcounter{equation}{0}

\renewcommand{\thefigure}{S\arabic{figure}}
\renewcommand\thesection{S\arabic{section}}
\renewcommand\thetable{S\arabic{table}}

\renewcommand{\vec}[1]{\boldsymbol{#1}}
\newcommand{\mat}[1]{\mathbf{#1}}

\newcommand{\revT}[1]{\textcolor{black}{#1}}

\newcommand{\alphafootnotemark}[1]{\textsuperscript{#1}}
\newcommand{\alphafootnotetext}[2]{\textsuperscript{#1}\footnotesize{#2}}

{\centering\Large{\textbf{Supporting Information: \\}}}
{\centering\Large{\textbf{Light-Induced Persistent Electronic Chirality in Achiral \\ Molecules Probed with Time-Resolved Electronic Circular Dichroism Spectroscopy}}}

{\noindent{Torsha Moitra, Lukas Konecny, Marius Kadek, Ofer Neufeld, Angel Rubio, Michal Repisky}}
\vspace{8em}

\subsection*{Table of Contents for Supporting Information}

\noindent
\textbf{S1. Methodology} \dotfill \pageref{sec:methods} \\
\textbf{S2. Computational setup} \dotfill \pageref{sec:comp} \\
\textbf{S3. Ground state depopulation} \dotfill \pageref{sec:gspop} \\
\textbf{S4. Analysis of induced electric dipole moment} \dotfill \pageref{sec:dipole} \\
\textbf{S5. Time evolution of induced charge and current density (video)} \dotfill \pageref{sec:video} \\
\textbf{S6. TR-ECD spectra of benzene and aniline} \dotfill \pageref{sec:others}

\clearpage
\section{Methodology}\label{sec:methods}

Let us start the theory section by considering the interaction of a molecule with an external monochromatic radiation field characterized by the electric $\vec{E}(\vec{r},t)$ and magnetic $\vec{B}(\vec{r},t)$ field components. The radiation field induces oscillating 
electric and magnetic multipole moments in a molecule. These moments are related to 
components of the radiation field through molecular property tensors. A key quantity for predicting the 
optical rotatory dispersion (ORD) or the electronic circular dichroism (ECD) spectra is the property tensor $\vec{\beta}$, which relates the induced electric ($\vec{\mu}$) and magnetic ($\vec{m}$) dipole moments to the time derivative of the fields (in SI-based atomic units)~\footnotemark[1]$^,$\footnotemark[2]:
\footnotetext[1]{D. Varsano, L. A. Espinosa-Leal, X. Andrade, M. A. Marques, R. Di Felice, and A. Rubio, Phys. Chem. Chem. Phys. 11, 4481 (2009).}
\footnotetext[2]{L. D. Barron, Molecular light scattering and optical activity (Cambridge University Press, 2009).} 
\begin{align}
  \label{eq:1}
  \mu_j(t) 
  & =
  \int_{-\infty}^{\infty}
  dt'~
  \alpha_{jk}(t-t')E_{k}(t')
  -
  \int_{-\infty}^{\infty}
  dt'~
  \beta_{jk}(t-t')\frac{\partial B_{k}(t')}{\partial t'}
  + ...
  \\[0.3cm]
  \label{eq:2}
  m_j(t) 
  & =
  \int_{-\infty}^{\infty}
  dt'~
  \chi_{jk}(t-t')B_{k}(t')
  +
  \int_{-\infty}^{\infty}
  dt'~
  \beta_{kj}(t-t')\frac{\partial E_{k}(t')}{\partial t'}
  + ...
\end{align}
Here, $\vec{\alpha}$ and $\vec{\chi}$ are the electric polarizability and magnetic susceptibility tensors, respectively, and summations are implicit for repeated indices denoting the Cartesian axes ($j, k$). Note that the first index of the tensor $\vec{\beta}$ connects to an electric quantity, while the second index to a magnetic quantity.

The previous relations ignore the nonlocal response in space. This approximation is valid when the wavelength of the optical waves of interest is long compared to the range of the considered response function. This approximation is appropriate for the present work where
the wavelength of visible light $\approx\!5000$ \AA~is much longer than the typical length scale of the organic molecules $\approx\!10\!-\!100$ \AA. Additional implication of this (long wavelength) approximation is that the electric field is spatially uniform and the magnetic field vanishes\footnotemark[2].
So, in the case of a polychromatic electric field given as
$$
  E_{k}(t) 
  =
  \int_{-\infty}^{\infty} 
  \frac{d\omega}{2\pi} E_{k}(\omega) e^{-i\omega t},
$$
the property tensor $\vec{\beta}$ can be obtained within the approximation from Eq.~\eqref{eq:2}
\begin{align}
  \label{eq:3}
  m_j(t) 
  & =
  \int_{-\infty}^{\infty}
  dt'~
  \Bigg[
  \int_{-\infty}^{\infty} 
  \frac{d\omega}{2\pi} \beta_{kj}(\omega) e^{-i\omega(t-t')}
  \Bigg]
  \Bigg[
  -i\int_{-\infty}^{\infty} 
  \frac{d\omega'}{2\pi} \omega' E_{k}(\omega') e^{-i\omega' t'}
  \Bigg]
  \nonumber
  \\
  & =
  \frac{-i}{2\pi}
  \int_{-\infty}^{\infty} 
  d\omega~\beta_{kj}(\omega) \omega E_{k}(\omega) e^{-i\omega t}.
\end{align}
Here, we used the identity 
$\int_{-\infty}^{\infty} dt~e^{i(\omega-\omega')t} = 2\pi\delta(\omega-\omega')$.
Now, as a particular case we consider an electric field with equal
intensity $(\kappa)$ for all frequencies and directions, \textit{i.e.}, 
$E_{k}(\omega) = \kappa_{k}$. In the time domain this corresponds to 
the Dirac delta-type field $E_{k}(t) = \kappa_{k}\delta(t)$. 
By introducing the expression for $E_{k}(\omega)$ into the previous equation, and performing an inverse Fourier transform, we obtain
\begin{align}
  \label{eq:4}
  \int_{-\infty}^{\infty}
  dt~m_j(t) e^{i\omega t}
  & =
  \frac{-i}{2\pi}
  \int_{-\infty}^{\infty}
  dt~ e^{i\omega t}
  \int_{-\infty}^{\infty} 
  d\omega'~\beta_{kj}(\omega') \omega' E_{k}(\omega') e^{-i\omega' t}
  \nonumber
  \\[0.2cm]
  & =
  -i\omega\kappa_{k} \beta_{kj}(\omega).
\end{align}
In other words, we see that the frequency-dependent property tensor 
$\vec{\beta}$ can be obtained from the time-dependent electrically induced magnetic dipole moment
\begin{align}
  \label{eq:5}
  \beta_{kj}(\omega)
  & =
  \frac{i}{\omega\kappa_{k}}  
  \int_{-\infty}^{\infty}
  dt~m_j(t) e^{i\omega t}.
\end{align}
The final ECD spectra are obtained from the imaginary part of the isotropically averaged $\vec{\beta}$:
$$
   \bar{\beta}(\omega) = \frac{1}{3}\sum_{k} \beta_{kk}(\omega),
$$
and reported in this work in terms of the differential extinction coefficient $\Delta\varepsilon$ in L/(mol$\cdot$cm)~\footnotemark[2]$^,$\footnotemark[3]:
\footnotetext[3]{L. Konecny, M. Repisky, K. Ruud, and S. Komorovsky, J. Chem. Phys. 151, 194112 (2019).} 
\begin{align}
 \nonumber
 \Delta\varepsilon(\omega)
 & =
 10\frac{2N_{A}\omega^{2}}{\ln(10)c^{2}\epsilon_{0}}\Im\Big[\bar{\beta}(\omega)\Big]
 \nonumber
 \\
 & =
 10\frac{2N_{A}}{\ln(10)c^{2}\epsilon_{0}} \omega \Im\Big[\frac{i}{3}\sum_{k}\frac{m_{k}(\omega)}{\kappa_{k}}\Big]
 \nonumber
 \\
 & =
 10\frac{2N_{A}}{\ln(10)c^{2}\epsilon_{0}} \frac{e^{2}\cdot a_{0}}{m_\text{e}} \omega^{\text{AU}}
 \Im\Big[\frac{i}{3}\sum_{k}\frac{m_{k}^{\text{AU}}(\omega)}{\kappa_{k}^{\text{AU}}}\Big].
 \label{eq:6}
\end{align}
Here, $N_A$ is the Avogadro constant, $c$ is the speed of light in vacuum, 
$\epsilon_{0}$ is the vacuum permittivity, $a_0$ is the Bohr radius, $m_e$ is the electron mass, and $e$ is the electron charge. All these physical quantities are expressed in SI units, except those with the superscript AU, which are expressed in SI-based atomic units (as obtained from our program's outputs).

The time-dependent induced magnetic dipole moment in Eq.~\eqref{eq:5} is calculated within the Kohn--Sham DFT framework as a trace of the magnetic dipole moment matrix $\mat{M}_{j}$ with the time-dependent density matrix $\mat{D}(t)$:
\begin{equation}
   \label{SIeq:7}
   m_{j}(t) 
   = 
   \Tr\!\big\{ \mat{M}_{j}\mat{D}(t) \big\} 
   - 
   m_{j}^{\text{static}}. 
\end{equation}
Here, $m_{j}^{\text{static}}$ refers to the static magnetic dipole moment obtained from the ground-state self-consistent field (SCF) density matrix $\mat{D}_{0}$ as: 
$m_{j}^{\text{static}} = \Tr\!\big\{ \mat{M}_{j}\mat{D}_{0} \big\}$. 
The matrix $\mat{M}_{j}$ consists of a finite-basis representation of the magnetic dipole
moment operator ($\hat{\vec{m}}$), namely its orbital angular momentum part,
\begin{equation}
    \label{SIeq:8}
    M_{j,ab} 
    =
    -\frac{1}{2}\bra{\chi_{a}} (\vec{r}\times\vec{p})_{j} \ket{\chi_{b}}.
\end{equation}
Since the present work restricts to closed-shell molecules, the spin angular momentum contribution to 
$\hat{\vec{m}}$ is negligible. The density matrix $\mat{D}(t)$ is obtained by solving the Liouville-von Neumann (LvN) equation of motion
\begin{equation}
	i \frac{\partial \mathbf{D}(t)}{\partial t} 
    = 
    [\mathbf{F}(t),  \mathbf{D}(t)],
\end{equation}
where the time evolution of a system characterized by 
$\mat{D}(t)$ is driven by the Fock matrix $\mathbf{F}(t)$. The Fock matrix describes the molecular system itself as well its interaction with external field(s), and it consists of one-electron ($\mathbf{h}$), two-electron ($\mathbf{G}$), exchange--correlation
($\mathbf{V}^{\mathrm{XC}}$), and light--matter interaction ($\mathbf{V}^{\mathrm{ext}}$) contributions:
\begin{equation}
    \label{eq:10}
	\mathbf{F}(t) 
    = 
    \mathbf{h} + \mathbf{G}[\mathbf{D}(t)] + \mathbf{V}^{\mathrm{XC}}[\mathbf{D}(t)] 
    + \mathbf{V}^{\mathrm{ext}}(t).
\end{equation}
All these matrices are represented in the basis of time-independent molecular orbitals (MOs) obtained from the solution of Kohn--Sham SCF equations, where each MO is composed of a linear combination of Gaussian-type orbitals (GTOs). In this work, we employed uncontracted aug-cc-pVTZ (for furan) and aug-cc-pVDZ (for aniline and benzene) GTOs~\footnotemark[4]$^,$\footnotemark[5].
\footnotetext[4]{T. H. Dunning Jr, J. Chem. Phys. 90, 1007 (1989).} 
\footnotetext[5]{R. A. Kendall, T. H. Dunning, and R. J. Harrison, J. Chem. Phys. 96, 6796 (1992).} 
We employ the PBE0 exchange–correlation functional for furan, and the PBE functional for aniline and benzene~\footnotemark[6]$^,$\footnotemark[7].
\footnotetext[6]{C. Adamo and V. Barone, J. Chem. Phys. 110, 6158 (1999).}
\footnotetext[7]{J. P. Perdew, K. Burke, and M. Ernzerhof, Phys. Rev. Lett. 77, 3865 (1996).}
All implementation and numerical simulations pertaining to this work were performed using the ReSpect program~\footnotemark[8].
\footnotetext[8]{M. Repisky, S. Komorovsky, M. Kadek, L. Konecny, U. Ekström, E. Malkin, M. Kaupp, K. Ruud, O. L. Malkina, and V. G. Malkin, J. Chem. Phys. 152, 184101 (2020).}

As discussed above, molecular chiroptical effects can be attributed to the oscillating magnetic dipole moment induced by an external radiation electric field. The interaction of this external field with a molecular system is incorporated into our approach through the light--matter interaction term in Eq.~\eqref{eq:10}: 
\begin{align}
    \label{eq:11}
	\mathbf{V}^{\mathrm{ext}}(t) 
    &= 
    -\mathbf{P}_{k}\mathcal{E}^\text{L/R}_{k}(t)  -\mathbf{P}_{k}\mathcal{F}_{k}(t).
\end{align}
In this work, we assume a time-resolved pump-probe experiment in the non-overlapping regime, where the system, characterized by the electric dipole moment ($\mat{P}$), interacts with both the external electric pump ($\bm{\mathcal{E}^\text{L/R}}$) and probe ($\bm{\mathcal{F}}$) fields. The probe pulse is applied at the end or after the pump pulse. The pump pulse has the form of a \textit{chiral}, circularly polarized left (L) or right (R) Gaussian envelope function:
\begin{align}
	\label{eq:pump}
	\bm{\mathcal{E}}^{\text{L/R}} (t) 
    =
    \begin{cases}
       \mathcal{E}_0 \bm{e^\text{L/R}} \exp(\frac{-(t-t_0)^2}{2\sigma^2}) \qquad t \leq T
       \\
       0 \qquad~\qquad~\qquad~\qquad~\quad t>T,
    \end{cases}
\end{align}
centered at $t_0$ and characterized by the amplitude $\mathcal{E}_0$, standard deviation $\sigma$, and helicity vector
\begin{align}
\label{eq:eLR}
	\bm{e^\text{L/R}} 
    = 
    \cos\big(\omega_0(t-t_0)\big)\bm{x} 
    \mp 
    \sin\big(\omega_0(t-t_0)\big)\bm{y}, 
\end{align}
which oscillates with the carrier frequency $\omega_0$. 
As the Gaussian envelope function decays asymptotically, we impose a step function for $\bm{\mathcal{E}}^{\text{L/R}}(t)$ at $t>T$. The pump duration $T$ corresponds to the time at which the pump amplitude reaches the value $10^{-2}\times \mathcal{E}_0$, and is related to the standard deviation $\sigma$ as
\begin{align}
    T &= 2 \sigma  \left\lfloor \revT{{\sqrt{-2  \ln(0.01)}}} \right\rfloor 
    .
\end{align}
In accordance with the previous discussion, the probe pulse $\bm{\mathcal{F}}(t)$ is modeled as an analytical delta function with the amplitude $\mathcal{F}_0$, direction $\bm{f}$, and origin at $T+\tau$: 
\begin{align}
	\label{eq:probe}
	\bm{\mathcal{F}}(t) &= \mathcal{F}_0 \bm{f} \delta \big(t-(T+\tau)\big),
\end{align}
where $\tau$ denotes the time delay between the pump and probe pulses. The specific values used in our simulations for the pump-probe pulse setup and real-time propagation are provided in Table~\ref{tab:setup}. 
Finally, the electric dipole operator represented in a finite basis has the matrix form
\begin{equation}
    P_{k,ab} 
    =
    -\bra{\chi_{a}} r_{k} \ket{\chi_{b}},
\end{equation}
where $\vec{r}$ is the electronic position operator.

In the pump-probe setup, one aims to compute the absorption of the probe pulse by a molecule that is initially irradiated by a pump, either simultaneously or with a given delay~\footnotemark[9]$^,$\footnotemark[10].
\footnotetext[9]{U. De Giovannini, G. Brunetto, A. Castro, J. Walkenhorst, and A. Rubio, ChemPhysChem. 14, 1363 (2013).} 
\footnotetext[10]{T. Moitra, L. Konecny, M. Kadek, A. Rubio, and M. Repisky, J. Phys. Chem. Lett. 14, 1714 (2023).}
For pump-probe ECD, the goal is therefore to determine the response of the magnetic dipole moment with ($\bm{m}^{\mathrm{pp}}$) and without ($\bm{m}^{\mathrm{p}}$) the probe pulse. The difference ($\Delta\bm{m}^{\text{TR-ECD}}$) represents the excess of magnetization, which is responsible for the absorption of the probe. Using Eqs.~\eqref{SIeq:7} and~\eqref{SIeq:8}, this requires evaluating
\begin{align}
	\Delta m^{\text{TR-ECD}}_{j}(t) 
	= 
    m^{\mathrm{pp}}_{j}(t)  - m^{\mathrm{p}}_{j}(t) 
	= 
    \Tr\!\left\{ \mathbf{M}_{j} \big( \mathbf{D}^{\mathrm{pp}}(t) - \mathbf{D}^{\mathrm{p}}(t) \big) \right\},   
	\label{eq:delta_m}
\end{align}
where $\mathbf{D}^{\text{p}}(t)$ and $\mathbf{D}^{\text{pp}}(t)$ are time-dependent density matrices obtained from two distinct simulations: one with only the pump pulse (p) and one with both the pump and probe pulses (pp).
Finally, the vector $\Delta \vec{m}^{\text{TR-ECD}}(t)$ is transformed to the frequency domain using the discrete Fourier transformation
\begin{align}
	\label{eq:m-FFT}
    \Delta\bm{m}^{\text{TR-ECD}} (\omega_a) 
    = 
    \sum_{b=0}^{N-1} \Delta t  \Delta \bm{m}^{\text{TR-ECD}}(t_b)
	\exp(-\gamma t_b) \exp({i 2\pi \frac{ab}{N}}). 
\end{align}
In the above equation, $\gamma$ is the real-valued damping constant and the
frequency index $a$ runs from 0 to $(N-1)$ where $N$ is the number of
time-steps in the real-time simulation and $\omega_a =\frac{2\pi a}{\Delta t}$.
Note here, that for the Fourier transformation $t=0$ is set to the time of
application of probe pulse, \textit{i.e.}, $T+\tau$. 

The final TR-ECD spectra were obtained from three independent pump-probe simulations with the \revT{uni-directional pump field and} probe field oriented along three Cartesian directions. This allows us to construct full frequency-dependent property tensor
$\vec{\beta}$ in the sense of Eq.~\eqref{eq:5}. Its isotropically averaged imaginary part leads to the TR-ECD differential extinction coefficient as in Eq.~\eqref{eq:6}.

\section{Computational setup} \label{sec:comp}
\noindent
All real-time TDDFT simulations employed the theoretical methodology discussed in the previous section, and were performed with 0.2 au time-step ($\Delta t$) for 20000 steps, resulting in a total simulation time of 96.75 fs. 
\revT{The UV-vis spectra is obtained with a $\delta$-type pulse of amplitude 0.001 au.}
The details of the pump and probe pulse parameters for the time-resolved studies are summarized in Table~\ref{tab:setup}.
The experimental geometry was used for the furan, aniline and benzene molecules~\footnotemark[11]
as given in Table~S2.
\footnotetext[11]{ National institute of standards and technology (NIST) experimental geometry data, accessed: Jan 2024}

\begin{table*}[h]
	\centering
	\caption{Summary of pump and probe pulse parameters used in all simulations. Mathematical forms of these pulses are given in Eqs.~\eqref{eq:pump}-\eqref{eq:probe}, and lists as:
		carrier frequency ($\omega_0$), pump amplitude
		($\mathcal{E}_0$), pump center ($t_0$), pump duration ($T$),  standard deviation ($\sigma$), and probe amplitude $\mathcal{F}_0$.
		The pulse shapes are presented in Figures~\ref{fig:SI-pulseshape-furan}-\ref{fig:SI-pulseshape-benzene}. }
        \begin{tabular}{@{}c@{}|@{}cc@{}|@{}cc@{}|@{}cc@{}|@{}cc@{}|@{}cc@{}|@{}cc@{}}
		\hline
        Molecule &
        \multicolumn{10}{c|}{$\bm{\mathcal{E}}^{\text{L/R}} (t)$} &
        \multicolumn{2}{c}{$\bm{\mathcal{F}} (t)$}\\
        \hline
		&  \multicolumn{2}{c|}{$\omega_0$} & \multicolumn{2}{c|}{$\mathcal{E}_0$} & \multicolumn{2}{c|}{$t_0$} & \multicolumn{2}{c|}{$T$} & \multicolumn{2}{c|}{$\sigma$} 
          & \multicolumn{2}{c}{$\mathcal{F}_0$} \\
		& au & eV & au &V/m & au & fs & au &fs & au &fs & au &V/m \\
		\hline
	Furan &  0.223 & 6.07 & 0.03 & $1.54 \times 10^{10}$ \alphafootnotemark{a} & 84.6 & 2.05& 169.2 & 4.09 & 28.2 & 0.68 & 0.001 & $5.14 \times 10^{8}$ \alphafootnotemark{b}\\
		Aniline & 0.236\alphafootnotemark{c} & 6.41 & 0.03 & $1.54 \times 10^{10}$ \alphafootnotemark{a} & 79.8 & 1.93 & 159.6 & 3.86 & 26.60 & 0.64  & 0.001 & $5.14 \times 10^{8}$ \alphafootnotemark{b}\\
		Benzene &  0.251 & 6.84 & 0.03 & $1.54 \times 10^{10}$ \alphafootnotemark{a} & 75.0 & 1.81 & 150.0 &3.63 & 25.0 & 0.60  & 0.001 & $5.14 \times 10^{8}$ \alphafootnotemark{b}\\
		\hline
	\end{tabular}
	\label{tab:setup}
\end{table*}
{\noindent\alphafootnotetext{a}{$I_0 = 3.16 \times 10^{13}\text{~W/cm}^2$;}}
{\noindent\alphafootnotetext{b}{$I_0 = 3.51 \times 10^{10}\text{~W/cm}^2$;}}
{\noindent\alphafootnotetext{c}{Chosen as it has the maximum intensity and is similar in energy with C$_6$H$_6$.}}

\begin{table}[h]
\centering
\label{tab:Geometry}
\caption{Cartesian coordinates of the molecules (in \AA)}
\begin{tabular}{cccc|cccc| cccc}
	\hline \hline
	\multicolumn{4}{c|}{Furan} & \multicolumn{4}{c|}{Benzene} & \multicolumn{4}{c}{Aniline} \\
    \hline
         & X & Y& Z & & X & Y& Z & & X & Y& Z \\
         \hline
         O  &  0.0000  &   0.0000  &   1.1626 & H &  2.4750 &  0.0000 & 0.0000 &      C      &     -0.2201 &  -1.1999 &  -0.0049 \\ 
         C  &  0.0000  &   1.0920  &   0.3487 & C &  1.3935 &  0.0000 & 0.0000 &      C      &      1.1643 &  -1.1947 &   0.0035 \\ 
         C  &  0.0000  &  -1.0920  &   0.3487 & C &  0.6967 &  1.2069 & 0.0000 &      C      &      1.8704 &   0.0000 &   0.0075 \\ 
         C  &  0.0000  &   0.7169  &  -0.9596 & H &  1.2375 &  2.1434 & 0.0000 &      H      &      1.6966 &  -2.1387 &   0.0075 \\ 
         C  &  0.0000  &  -0.7169  &  -0.9596 & C & -0.6967 &  1.2069 & 0.0000 &      C      &      1.1643 &   1.1947 &   0.0035 \\ 
         H  &  0.0000  &   2.0473  &   0.8439 & H & -1.2375 &  2.1434 & 0.0000 &      H      &      2.9528 &   0.0000 &   0.0143 \\ 
         H  &  0.0000  &  -2.0473  &   0.8439 & C & -1.3936 &  0.0000 & 0.0000 &      C      &     -0.2201 &   1.1999 &  -0.0049 \\ 
         H  &  0.0000  &   1.3509  &  -1.8290 & H & -2.4750 &  0.0000 & 0.0000 &      H      &      1.6966 &   2.1387 &   0.0075 \\ 
         H  &  0.0000  &  -1.3509  &  -1.8290 & C & -0.6968 & -1.2069 & 0.0000 &      C      &     -0.9339 &  -0.0000 &  -0.0084 \\ 
            &          &           &          & H & -1.2375 & -2.1434 & 0.0000 &      H      &     -0.7591 &   2.1414 &  -0.0130 \\ 
            &          &           &          & C &  0.6968 & -1.2069 & 0.0000 &      H      &     -0.7591 &  -2.1414 &  -0.0130 \\ 
            &          &           &          & H &  1.2375 & -2.1434 & 0.0000 &      N      &     -2.3199 &   0.0000 &  -0.0724 \\ 
            &          &           &          &   &           &           &    &      H      &     -2.7690 &   0.8355 &   0.2628 \\ 
            &          &           &          &   &           &           &    &      H      &     -2.7690 &  -0.8355 &   0.2628 \\ 
\hline \hline
\end{tabular}
\end{table}

\begin{table}[htb!]
    \centering
    \begin{tabular}{c|c|c}
    \hline\hline
    Molecule & Level & {IE $= - \epsilon^{\text{HOMO}}$} \\ \hline 
        Furan  & PBE0/aug-cc-pVTZ & 0.24559 au \\
        Aniline & PBE/aug-cc-pVDZ & 0.18236 au \\
        Benzene & PBE/aug-cc-pVDZ & 0.23240 au \\
        \hline\hline
    \end{tabular}
    \caption{\revT{Ionization energies from Koopman´s theorem calculated as the negative of the highest occupied KS orbital energy. Although more rigorous methodologies, like $\Delta$-SCF, are available, we do not pursue them as the energies are only a qualitative energy boundary and are not used for any analysis. }}
    \label{tab:IE}
\end{table}

\begin{figure}[h]
		\centering
	\includegraphics[width=0.7\textwidth]{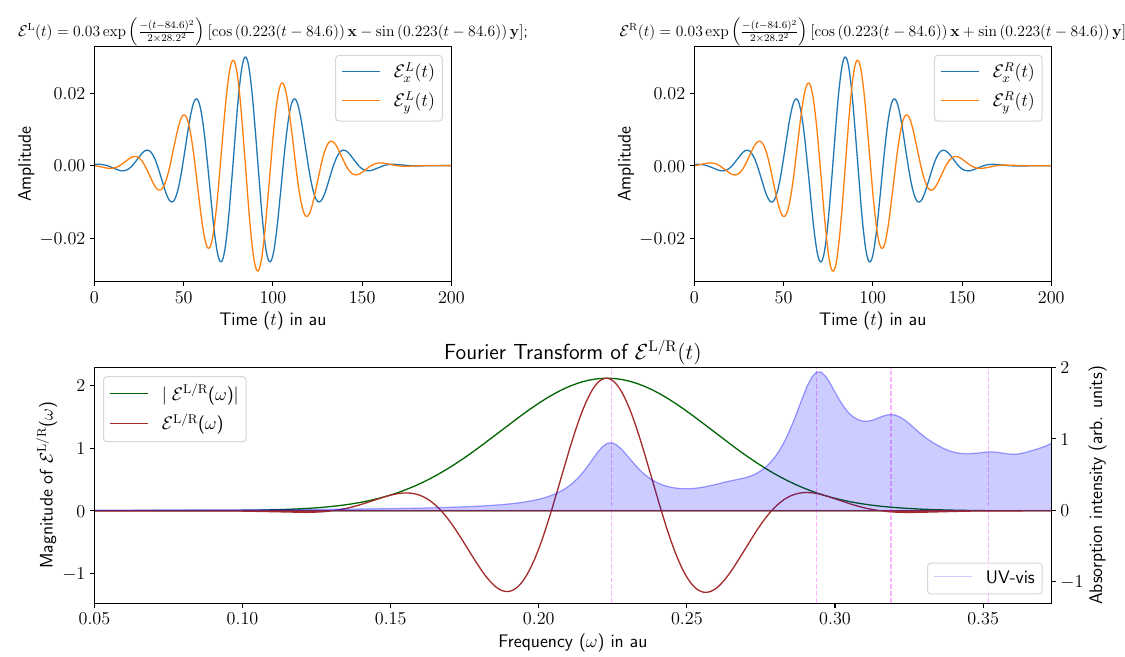}
\caption{Furan: CPL (top left) and CPR (top right) pump features in time (top) and frequency (bottom) domain. The ground state bright excitation energies are marked by magenta dashed lines. 
}
    \label{fig:SI-pulseshape-furan}
\end{figure}

\begin{figure}[H]
	\centering
	 \includegraphics[width=0.7\textwidth]{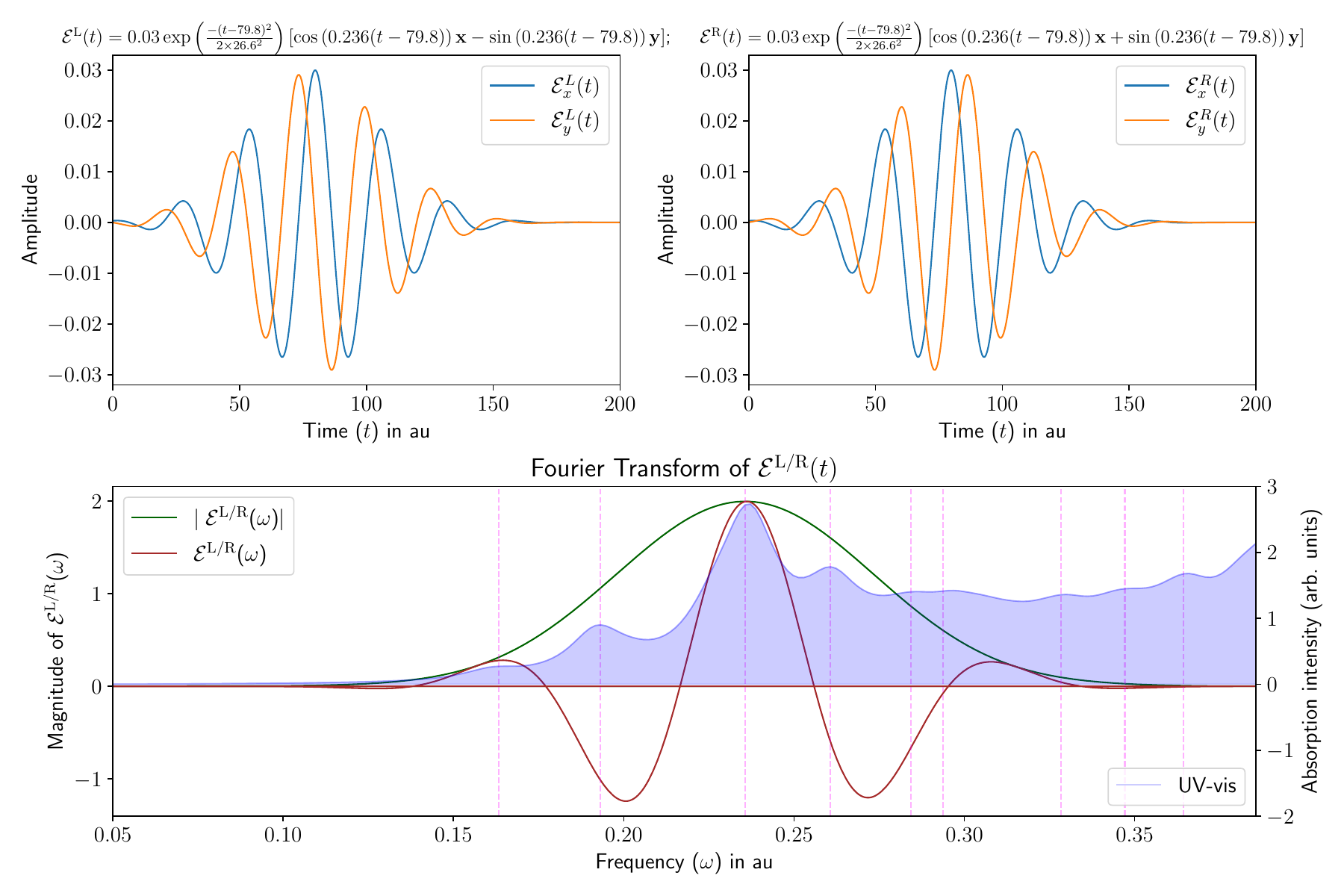}
	\caption{Aniline: CPL (top left) and CPR (top right) pump features in time (top) and frequency (bottom) domain. The ground state bright excitation energies are marked by magenta dashed lines.	
    }
\label{fig:SI-pulseshape-aniline}
\end{figure}

\begin{figure}[H]
		\centering
	\includegraphics[width=0.7\textwidth]{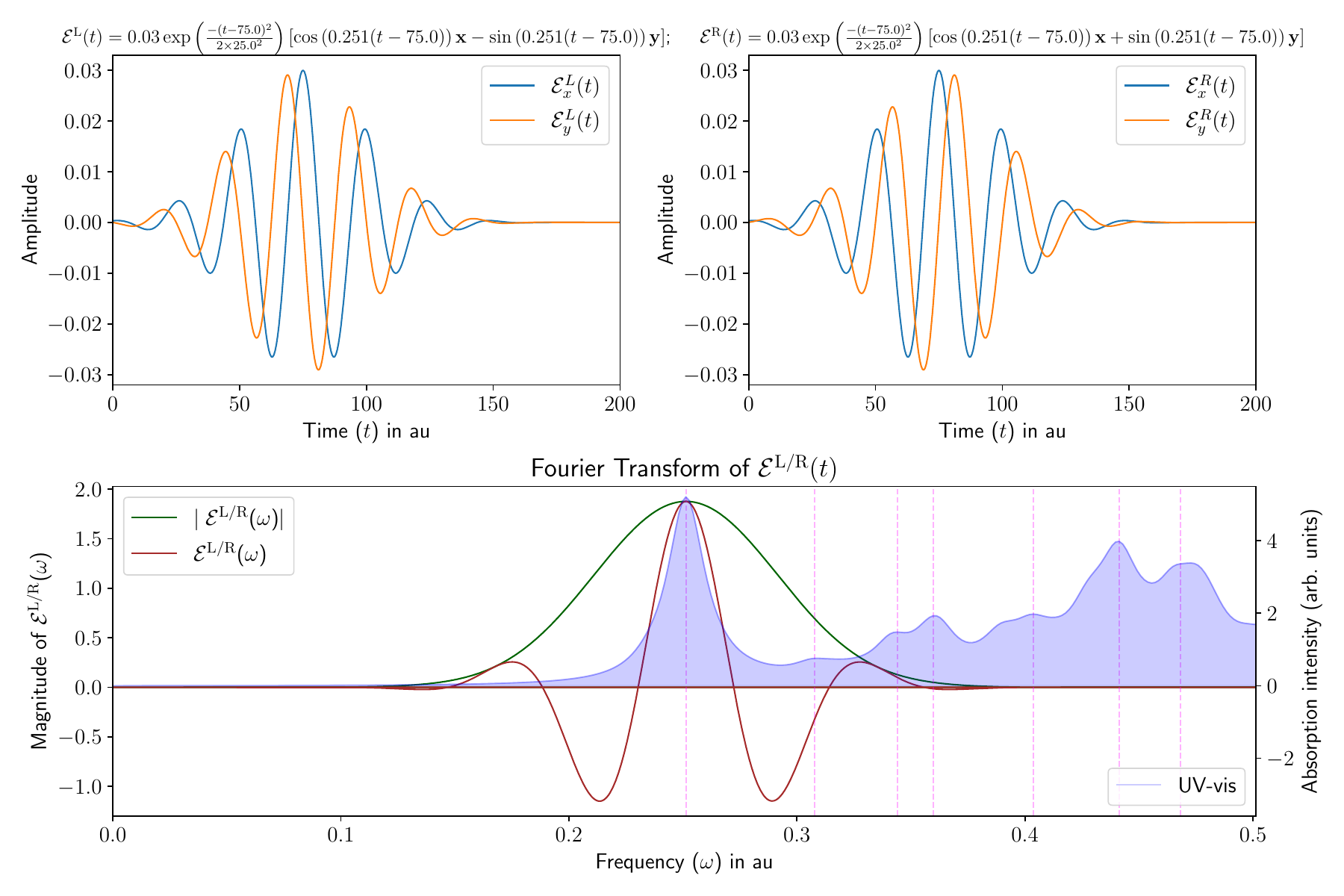}
	\caption{Benzene: CPL (top left) and CPR (top right) pump features in time (top) and frequency (bottom) domain. The ground state bright excitation energies are marked by magenta dashed lines. 
    }
    \label{fig:SI-pulseshape-benzene}
\end{figure}

In addition, both the time-domain and Fourier-transformed frequency-domain representations of the broadband circularly polarized left- and right-handed pump pulses are shown in Figs.~\ref{fig:SI-pulseshape-furan}-\ref{fig:SI-pulseshape-benzene}. The carrier frequency ($\omega_0$) of the pump pulse is tuned to the first excited state (unless otherwise mentioned), while the pulse duration is chosen so that its frequency bandwidth also encompasses the second bright excited state. This results in more intricate coherent dynamics within the molecule.


\section{Ground state depopulation} \label{sec:gspop}
\noindent
\begin{figure}[h]
    \centering
    \includegraphics[width=0.6\textwidth]{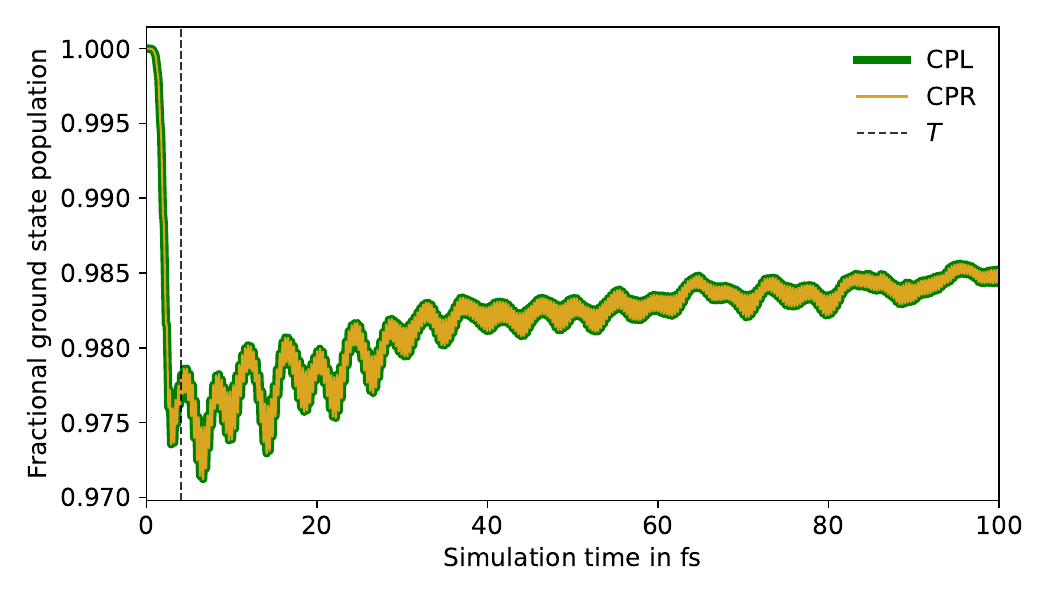}
    \caption{Furan: Variation in
        the fractional contribution of ground state towards the chiral electronic
        wavepacket computed as $\Tr[ \mathbf{D}_{0}\mathbf{D}(t)]$ with simulation
        time ($t$). $\mathbf{D}_{0}$ and $\mathbf{D}(t)$ are the reduced one-electron
        density of the ground state at time zero and non-stationary state at time $t$,
        respectively. }
    \label{fig:SI_gs_depop}
\end{figure}
We obtain the fractional ground state population as $\Tr[ \mathbf{D}_{0}\mathbf{D}(t)]$. 
For exact theory, the ground-state population after the end of pump and probe pulses should be constant.  
However, this does not hold for approximate mean-field methods such as HF or DFT, resulting in artificial oscillations in ground-state population, as seen in Fig.~\ref{fig:SI_gs_depop} and discussed in further details in Ref.\footnotemark[10]$^,$\footnotemark[12]$^,$\footnotemark[13].
\footnotetext[12]{J. I. Fuks, N. Helbig, I. Tokatly, and A. Rubio, Phys. Rev. B 84, 075107 (2011).}
\footnotetext[13]{J. I. Fuks, K. Luo, E. D. Sandoval, and N. T. Maitra, Phys. Rev. Lett. 114, 183002 (2015).}

As is well established in transient absorption spectroscopy, sufficient ground-state depopulation is crucial for generating an electronic wavepacket with distinct characteristics and a unique spectral signature~\footnotemark[10].
Fig.~\ref{fig:SI_gs_depop} shows that the pump pulse setup of our choice leads to a ground-state depopulation of approximately 3\%. 
\revT{Following the pump pulse, the electronic wavepacket undergoes coherent dephasing, giving the appearance of relaxation toward the ground-state density due to destructive interference among its eigenstate components.}

\section{Analysis of induced electric dipole moment}\label{sec:dipole}
\noindent
\begin{figure}[h]
    \centering
    \includegraphics[width=\linewidth]{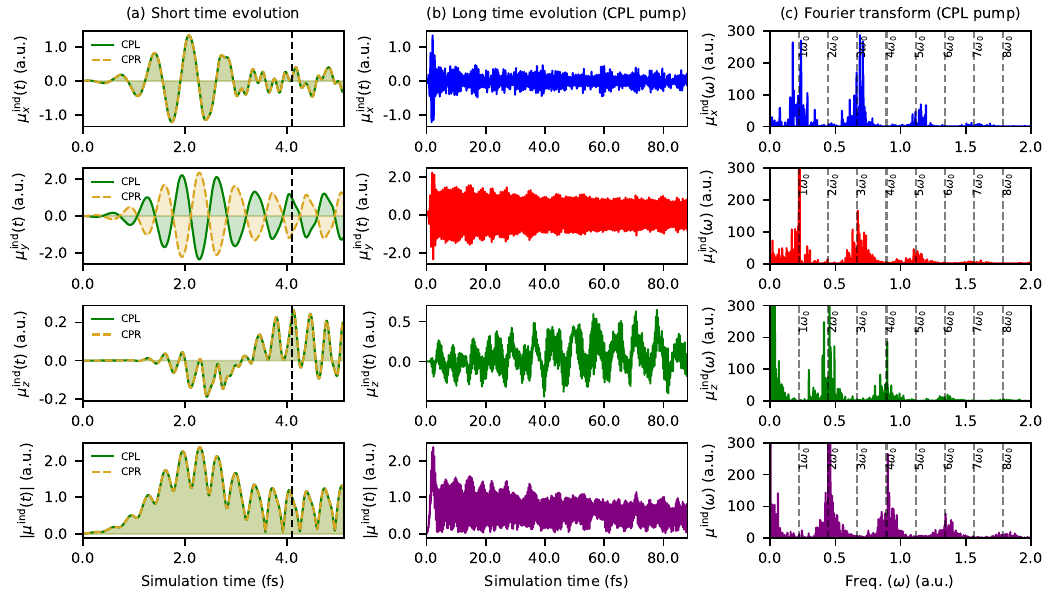}
    \caption{
Furan: 
Short time evolution (a) and long time evolution (b) of the magnitude ($|\bm{\mu}^{\text{ind}}(t)|$) and components of electric dipole moment $\mathbf{\mu}^{\text{ind}}(t)=(\mu^{\text{ind}}_x(t), \mu^{\text{ind}}_y(t), \mu^{\text{ind}}_z(t))$ induced by CPL and/or CPR pump pulses. The black dashed line in (a) at 4.09 fs marks the end of the pump pulse, after which the electronic wavepacket evolves freely.
    (c) Fourier transform of the induced electric moment obtained from the long time evolution in (b). 
    The harmonic orders of the carrier frequency $\omega_0=0.223$ au are marked by dashed lines.  
    }
    \label{fig:mu_t}
\end{figure}

\noindent
As discussed in the Letter, the oscillations in the induced magnetic moment (shown in Fig. 2) is not characterized by a single dominant frequency, instead it comprises of multiple harmonic orders of the carrier frequency ($\omega_0$). A similar behavior is observed for the induced electric dipole moment $\mu (t)$ signal as shown in Figure~\ref{fig:mu_t}, as is more commonly known for high harmonic generation (HHG)~\footnotemark[14]$^,$\footnotemark[15]$^,$\footnotemark[16]
\footnotetext[14]{O. Neufeld, D. Ayuso, P. Decleva, M. Y. Ivanov, O. Smirnova, and O. Cohen, Phys. Rev. X. 9, 031002 (2019). }
\footnotetext[15]{D. Ayuso, A. F. Ordonez, P. Decleva, M. Ivanov, and O. Smirnova, Opt. Express 30, 4659 (2022).} 
\footnotetext[16]{R. Cireasa, A. Boguslavskiy, B. Pons, M. Wong, D. Descamps, S. Petit, H. Ruf, N. Thiré, A. Ferré, J. Suarez, et al., Nat. Phys. 11, 654 (2015).}

\section{Time evolution of induced charge and current density (video)} \label{sec:video}
\noindent
A video showing the time evolution of the induced charge and current densities of furan are attached (Fig. SI\_charge\_current.mov). The primary observations are: (i) The enantiomer-like relationship between the electronic wavepackets induced by CPL and CPR is maintained both during and after the pump pulse. (ii) Notably, the induced chiral current density persists even after the pump pulse has ended, allowing its detection. 

\section{TR-ECD spectra of benzene and aniline} \label{sec:others}
\begin{figure*}[htb!]
\includegraphics[width=0.8\textwidth]{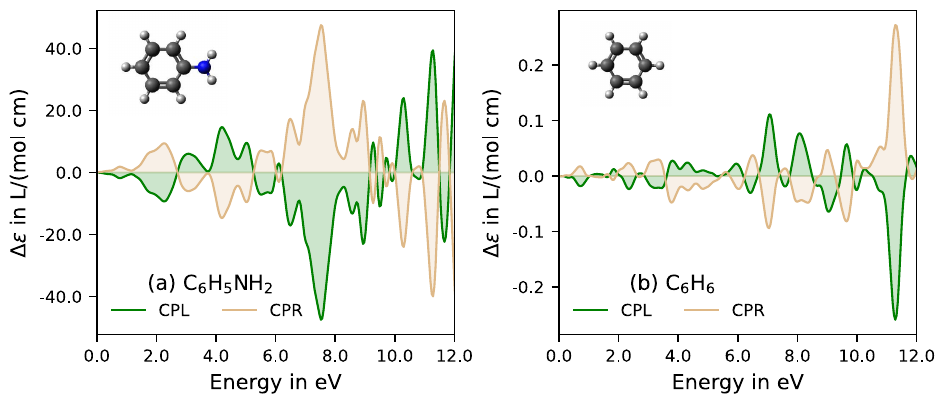}
\caption{Time-resolved electronic circular dichroism spectrum as differential extinction coefficient ($\Delta \varepsilon$) of (a) aniline (C$_6$H$_5$NH$_2$) and (b) benzene (C$_6$H$_6$) obtained at time-delay $\tau=0.0$. The pump-probe setup parameters are described in Table~\ref{tab:setup}.}
\label{fig:benzene-aniline-TRECD}
\end{figure*}
We generalize the proposed technique by further investigating planar achiral benzene and aniline molecules. 
\revT{Note that the relative orientation of the pump and molecule differs between the furan case (light propagates along $z$ within the molecular plane) and the aniline/benzene case (light propagates along $z$ perpendicular to the molecular plane). }
The TR-ECD spectra obtained for the molecules at time-delay $\tau=0.0$ fs is shown in  
Fig.~\ref{fig:benzene-aniline-TRECD}. We observe a mirror-image relationship between the CPL and CPR induced spectral function, concurrent with observations for furan molecule. 
However, our study on furan along with literature reports on
NaK~\footnotemark[17],
suggests a hypothesis that a
ground state static electric dipole moment may be necessary for observing induced
electronic chirality. Our simulations on benzene and aniline
reveal that while chirality is indeed induced in achiral benzene using a monochromatic
CP light pump pulse, 
the spectral intensity is two orders of magnitude smaller than that of aniline.
\footnotetext[17]{Y. Chen, D. Haase, J. Manz, H. Wang, and Y. Yang, Nat. Comm. 15, 565 (2024).}

\end{document}